\documentstyle[12pt]{article}
\begin{document}
\begin{titlepage}
\rightline{RIMS-948}
\rightline{October 1993}
\bigskip\bigskip
\begin{center}
{\Large Level two irreducible representations of $U_q(\widehat{sl}_2)$, \\
vertex operators, and their correlations}\\[12mm]
{\large Makoto Idzumi$^{1,2}$}\\[3mm]
Research Institute for Mathematical Sciences \\
Kyoto University, Kyoto 606, Japan \\[27mm]
\end{center}
\begin{abstract}
Vertex operators associated with level two $U_q(\widehat{sl}_2)$ modules
are constructed explicitly using bosons and fermions.
An integral formula is derived for the trace of products of vertex operators.
These results are applied to give $n$-point spin correlation functions
of an integrable $S=1$ quantum spin chain, extending an earlier work of
Jimbo {\em et al} for the case $S=1/2$.
\end{abstract}
\footnotetext[1]{
Permanent address: Department of Applied Physics, Faculty of Engineering,
University of Tokyo, Tokyo 113, Japan
}
\footnotetext[2]{
e-mail: {\tt idzumi@kurims.kyoto-u.ac.jp}
}
\end{titlepage}
\def\bF{{\bf F}}
\def\Bigbr#1{\Big\langle#1\Big\rangle}
\def\Bigp#1{\Big(#1\Big)_\infty}
\def\bbr#1{\langle\langle#1\rangle\rangle}
\def\br#1{\langle#1\rangle}
\def\bra#1{\langle#1|}
\def\C{{\bf C}}
\def\c{\angle}
\def\ceil#1{\lceil#1\rceil}
\def\ep{\epsilon}
\def\F{{\cal F}}
\def\floor#1{\lfloor#1\rfloor}
\def\head#1^#2{\buildrel {\scriptscriptstyle #2} \over #1}
\def\id{{\rm id}}
\def\ket#1{|#1\rangle}
\def\La{\Lambda}
\def\oi#1{\oint{dw_#1\over 2\pi i}}
\def\ov{\over}
\def\p#1{(#1)_\infty}
\def\Pf{{\rm Pf}\;}
\def\sid{{\rm{\scriptstyle id}}}
\def\tr{{\rm tr}\;}
\def\u{U_q(\widehat{sl}_2)}
\def\Up{U'_q(\widehat{sl}_2)}
\def\Z{{\bf Z}}

\section{Introduction}
Since Drinfeld \cite{Drinfeld1} and Jimbo \cite{Jimbo} introduced
quantum groups,  which are one parameter deformation of
the universal enveloping algebras of Kac-Moody Lie algebras
with the Hopf algebra structure,
it has been recognized that,
besides purely theoretical interests from mathematics,
the quantum groups have many applications and play important roles
in various fields in mathematics and physics.

Finding out explicit constructions of fundamental representations
of the algebra is a natural and important subject in the study.
Level one irreducible highest weight modules were constructed by
Frenkel {\em et al} \cite{FrJ} for simply laced quantum affine algebras,
the construction being called bosonization;
the structure of the modules was entirely analogous to the
Frenkel-Kac construction for affine Lie algebras.
Constructions of irreducible modules of higher levels are, however,
quite involved and there is no general recipe.
An exception is the construction via Wakimoto modules
for generic complex value of level \cite{Matsuo,KQS};
but it seems that the construction for integral level is not complete yet,
in which one needs to take quotient to get the irreducible one.
Bernard \cite{Bernard} found construction of
level one modules over $U_q(B_r^{(1)})$;
his results for $U_q(B_1^{(1)})$ can be translated into those for
level two $U_q(\widehat{sl}_2)$-modules.
One of the aims of this paper is to recall a part of his results and
to give explicit constructions
of level two irreducible highest weight modules over
the quantum affine algebra $\u$, which we denote
$V(2\Lambda_0)$, $V(2\Lambda_1)$ and $V(\Lambda_0+\Lambda_1)$
($\Lambda_i$ being the fundamental weights of $\widehat{sl}_2$);
the modules are constructed in terms of a single boson,
a fermion of Neveu-Schwarz or Ramond type
(according to the highest weights
$2\Lambda_{0,1}$ or $\Lambda_0+\Lambda_1$, respectively),
and the weight lattice of $sl_2$.
The result is analogous to that for $\widehat{sl}_2$ by
Lepowsky {\em et al} \cite{Lepowsky-Primc}.

Our second aim is to give explicit forms of $q$-deformed vertex operators
of type~I (in the sense of Ref.\cite{DFJMN})
associated with the level two modules defined as
homomorphisms of $\u$-modules
\[
\Phi(z): V(\lambda_m)\longrightarrow V(\lambda_{2-m})\otimes V_z
\]
where $\lambda_m=(2-m)\Lambda_0+m\Lambda_1,m=0,1,2$,
$V_z=V\otimes{\bf C}[z,z^{-1}]$, $V\simeq\C^3$.
In conformal field theories (CFTs)
vertex operators of this type
associated with the affine Lie algebras
are called primary fields,
and their $n$-point functions
\begin{equation}
\bra{\lambda}\Phi(z_1)\cdots\Phi(z_n)\ket{\lambda}
\quad (\ket{\lambda}: \hbox{highest weight vector})
\label{**}
\end{equation}
describe correlations of primary fields.
The $n$-point functions (\ref{**}) satisfy differential equations
called the Knizhnik-Zamolodchikov (KZ) equations.
Analogously,
Frenkel {\em et al} \cite{FrR} considered the $n$-point functions (\ref{**})
for the $q$-deformed vertex operators,
and showed that (\ref{**})  satisfy difference equations
called the $q$-deformed Knizhnik-Zamolodchikov ($q$-KZ) equations.
We generalize (\ref{**}) and consider
\begin{equation}
\tr_{V(\lambda)}(\xi^{-2d}y^\alpha\Phi_{i_1}(z_1)\cdots\Phi_{i_n}(z_n))
\label{*}
\end{equation}
for the vertex operators
associated with level two irreducible modules over $\u$,
where $\xi$ and $y$ are free parameters, $d$ the grading operator in $\u$,
and $\alpha$ the simple root of $sl_2$;
the previous ones (\ref{**}) are obtained from ours (\ref{*})
by specializing $\xi\to 0$ as coefficients of appropriate powers of $y$.
We note that
our $n$-point functions (\ref{*}), too, satisfy $q$-difference equations
of $q$-KZ type
(cf. Refs.\cite{IIJMNT,JMN,FJMMN};
we will not study the difference equations in this paper).
The goal of this paper is to give an integral formula
for the $n$-point functions (\ref{*});
the final formula is eq.(\ref{integral-formula-0}) or
eq.(\ref{integral-formula-1}).
We shall observe that
the integral formula gets simplified
at a special value of the parameter $\xi=q^2$,
which is due to delta functions arising from the fermion trace;
this special case is important because
the formula just at $\xi=q^2$ has an application
in the study of correlations in integrable spin systems and lattice models.

In recent progress in the study of integrable quantum spin systems and
lattice models
it has been recognized that
the $q$-deformed vertex operators play important roles in the models
that have quantum affine algebra symmetry \cite{DFJMN,JMMN,IIJMNT}.
Davies {\em et al} \cite{DFJMN} recognized that the XXZ model,
a spin~$1/2$ quantum spin chain, has an exact symmetry
$[H,\Up]=0$,
$H$ being the Hamiltonian, $\Up$ a subalgebra of $\u$
without the grading operator~$d$;
they conjectured that the space of states is a level zero $\u$~module
$\F \equiv \bigoplus_{\lambda,\lambda'}V(\lambda)\otimes V(\lambda')^{*a}
\simeq
\bigoplus_{\lambda,\lambda'} {\rm Hom}_\C(V(\lambda), V(\lambda))$,
$\lambda,\lambda'$ being level one highest weights in their case;
a physical state in $(\C^2)^{\otimes\infty}$ is identified with
a vector in $\F$ via $q$-deformed vertex operators;
they succeeded to construct the creation/annihilation operators in the model
and diagonalize the Hamiltonian.
Following this work
Jimbo {\em et al} \cite{JMMN} showed that
an arbitrary zero-temperature spin correlation function of the
XXZ model is formulated as
a trace of the product of vertex operators over
an irreducible highest weight $\u$-module of level one.
They obtained an integral formula for the $n$-point function
using the Frenkel-Jing construction of level one modules \cite{FrJ}.
Extensions to higher levels (spins) and incorporations of the formulation
in \cite{DFJMN} were done
by Idzumi {\em et al} \cite{IIJMNT}.

A motivation of the present paper was
to extend the work of Jimbo {\em et al} \cite{JMMN} for the case $S=1/2$
to $S=1$.
We will apply our integral formula for the level~two
vertex-operator $2n$-point functions
to zero-temperature spin $n$-point correlation functions
of an integrable $S=1$ spin chain
which is an integrable extension of the XXZ model to spin one:
$H = \sum_{l\in\Z} \cdots\otimes 1\otimes h\otimes 1\otimes\cdots$, where
\def\ch{{\rm ch}}
\def\sh{{\rm sh}}
\begin{eqnarray}
h&=&
 s^x\otimes s^x +  s^y\otimes s^y +  \ch(2\eta) \cdot s^z\otimes s^z -
 \Bigl(\sum_{j=1}^3 s^j\otimes s^j \Bigr)^2 \nonumber \\
&&+ 2\sh^2(\eta) \cdot
 \Bigl[ (s^z)^2\otimes \id + \id\otimes (s^z)^2 -
 (s^z\otimes s^z)^2 - 2\cdot\id\otimes\id  \Bigr] \nonumber \\
&&+ (2+4e^{-\eta}\ch^2(\eta)) \cdot (s^x\otimes s^x + s^y\otimes s^y)
s^z\otimes s^z \nonumber \\
&&+ (2+e^{\eta}) \cdot  s^z\otimes s^z (s^x\otimes s^x + s^y\otimes s^y);
\label{Ham}
\end{eqnarray}
here we have set $q=-e^{-\eta}$.    
We note that they can be regarded as the thermal average of
a product of in-row $n$~variables
for an integrable nineteen-vertex model,
whose Boltzmann weights being given by
\begin{equation}
\check R(z,w)=
\left[ \matrix{
1 & & & & & & & & \cr
 & e_1& & p& & & & & \cr
 & & c_1& & g_1& & b& & \cr
 & p& & e_2& & & & & \cr
 & & h_1& & o& & h_2& & \cr
 & & & & & e_1& & p& \cr
 & & b& & g_2& & c_2& & \cr
 & & & & & p& & e_2& \cr
 & & & & & & & & 1 \cr
} \right]
\label{R}
\end{equation}
where
$b = q^2(w-z)(q^2w-z)/d_2d_4$,
$c_1 = z^2(1-q^2)(1-q^4)/d_2d_4$,
$c_2 = w^2(1-q^2)(1-q^4)/d_2d_4$,
$e_1 = z(1-q^4)/d_4$,
$e_2 = w(1-q^4)/d_4$,
$p = q^2(w-z)/d_4$,
$g_1 = z(w-z)(q^2w-z)q(1-q^2)/d_2d_4$,
$g_2 = w(w-z)(q^2w-z)q(1-q^2)/d_2d_4$,
$h_1 = z(w-z)(q^2w-z)q(1+q^2)(1-q^4)/d_2d_4$,
$h_2 = w(w-z)(q^2w-z)q(1+q^2)(1-q^4)/d_2d_4$,
$o = q^2(w-z)(w-q^2z)+(1-q^2)(1-q^4)zw/d_2d_4$
with
$d_2 = w-zq^2$, $d_4 = w-zq^4$.

Finally we give a remark.
The vertex-operator correlation functions (\ref{*})
for general level can be calculated
through the Wakimoto constructions of highest weight modules
which are not irreducible by themselves;
cf. \cite{Matsuo,KQS}.
The final formula includes the Jackson-type integrals.
Compared with their approach through the Wakimoto constructions,
our bosonizations are more direct and
our final formula
(\ref{integral-formula-0}) or (\ref{integral-formula-1})
for (\ref{*})
gives rich information on the properties of the correlations (\ref{*}),
although we are restricted to level two only.

The paper is organized as follows.
In Section~\ref{sec-defs} we give several definitions needed in the
subsequent sections:
the quantum affine algebra $\u$,
the Drinfeld realization of the subalgebra $\Up$,
irreducible highest weight modules of level~$k$,
associated $q$-deformed vertex operators,
their $n$-point functions (\ref{*}),
and spin correlation functions for the spin-$k/2$ analog of the XXZ model;
there we will not specialize the level and we will give definitions
for general positive integer~$k$.
{}From Section~\ref{sec-modules} to the end we consider $k=2$ only.
In Section~\ref{sec-modules} we give explicit constructions (bosonizations)
of level~$k=2$ irreducible highest weight modules over $\u$;
besides a boson contained in the Drinfeld realization of $\Up$,
we need a Neveu-Schwarz or Ramond fermion as well as the weight lattice
of $sl_2$.
In Section~\ref{sec-VO} we present explicit forms of vertex operators (VOs)
(Eq.(\ref{VO})) and
prove intertwining relations with Chevalley generators.
In Section~\ref{sec-formula}
we give an integral formula for the VO $n$-point function;
it is an $n$-multiple contour integral of a meromorphic function
(Eqs.(\ref{integral-formula-0}) and (\ref{integral-formula-1}):
the main result of this paper).
We illustrate it for the VO two-point functions in
Section~\ref{sec-formula-example}.
In Section~\ref{sec-simplified}
the simplification at $\xi=q^2$ mentioned earlier is explained.
As an example
we show all VO two-point functions simplified at $\xi=q^2$
(Section \ref{sec-simplified-example}).
Finally we give an application to the spin correlation functions
for the spin-$1$ analog of the XXZ model,
which corresponds to a further specialization of the general formula
(Section \ref{sec-simplified-spin}).
In Appendix we assemble some formulae concerning boson fermion calculus
needed for the proof in Section~\ref{sec-VO} and
calculations in Section~\ref{sec-formula}.

\section{Definitions for general level $k$}
\label{sec-defs}
We give here definitions needed in this paper, which are
the algebra $\u$, irreducible highest weight modules, vertex operators
and their $n$-point functions.
Spin correlation functions are also given.
We shall follow the notations of Ref.\cite{IIJMNT}.

In this section we do not specialize a level $k \in\Z_{\ge 0}$.
In the subsequent sections where we shall give explicit constructions of
modules and vertex operators,
we shall fix the level $k=2$.

Throughout this paper we assume $q^n\not=1$ for any $n\in\Z_{\not=0}$.
In an application to a spin chain or a vertex model
we further assume $|q|<1$.

\subsection{Quantum affine algebra $\u$}
We give a definition of the quantum affine algebra $\u$
which is a deformation with a parameter $q$ of
the universal enveloping algebra of an affine Lie algebra $\widehat{sl}_2$.

We first fix the notations for the affine Lie algebra $\widehat{sl}_2$.
Let $\Lambda_0$, $\Lambda_1$ be the fundamental weights and
let $\delta$ be the null root.
They span the weight lattice $P=\Z\Lambda_0\oplus\Z\Lambda_1\oplus\Z\delta$.
The simple roots are given by
$\alpha_1=2\Lambda_1-2\Lambda_0$, $\alpha_0=\delta-\alpha_1$.
We normalize the invariant symmetric bilinear form on $P$ by
$(\alpha_i,\alpha_i)=2$.
Let $P^*$ be the dual lattice to $P$ and let
$\{h_0,h_1,d\}$ be the basis of $P^*$ dual to $\{\Lambda_0,\Lambda_1,\delta\}$
with respect to the natural pairing $\br{\quad,\quad} : P\times P^* \to \Z$.
Therefore
\[
(h_i,h_j)=\left\{\begin{array}{clr}
                   2 & (i=j) \\
                  -2 & (i\not= j)
                 \end{array} \right. ,\quad
(h_i,d)=\delta_{i,0},\quad
(d,d)=0;
\]
\begin{equation}
(\Lambda_i,\Lambda_j)={1\ov 2}\delta_{i,1}\delta_{j,1},\quad
(\Lambda_i,\delta)=1,\quad
(\delta,\delta)=0,
\label{inner-product}
\end{equation}
\[
(\alpha_i,\alpha_j)=\left\{\begin{array}{clr}
                   2 & (i=j) \\
                  -2 & (i\not= j)
                 \end{array} \right. ,\quad
(\alpha_i,\Lambda_0)=\delta_{i,0}.
\]
We use the standard notation $\rho=\Lambda_0+\Lambda_1$.

The quantum affine algebra $\u$ is an associative algebra over a field
$F={\bf C}$ with unit $1$ generated by
$e_i,f_i(i=0,1)$, $q^h(h\in P^*)$ with relations
\[
q^hq^{h'}=q^{h+h'},\quad q^0=1,
\]
\[
q^he_iq^{-h}=q^{\br{\alpha_i,h}}e_i,\quad
q^hf_iq^{-h}=q^{-\br{\alpha_i,h}}f_i,
\]
\[
e_if_j-f_je_i=\delta_{ij}{t_i-t_i^{-1}\ov q-q^{-1}}\quad(t_i=q^{h_i}),
\]
\[
e_i^3e_j-[3]e_i^2e_je_i+[3]e_ie_je_i^2-e_je_i^3=0\quad(i\not= j),
\]
\[
f_i^3f_j-[3]f_i^2f_jf_i+[3]f_if_jf_i^2-f_jf_i^3=0\quad(i\not= j)
\]
where $h,h'\in P^*$, $i,j=0,1$.
We use the notations
\[
[x]={q^x-q^{-x}\ov q-q^{-1}},\quad [n]!=[n][n-1]\cdots[2][1],\quad
\left[ \begin{array}{c}n\\ k\end{array} \right]
={[n]!\ov[k]![n-k]!}.
\]
The coproduct and the antipode are given by the formulae
\[
\Delta(e_i)=e_i\otimes 1+t_i\otimes e_i,\quad
\Delta(f_i)=f_i\otimes t_i^{-1}+1\otimes f_i,\quad
\Delta(q^h)=q^h\otimes q^h,
\]
\[
a(e_i)=-t_i^{-1}e_i,\quad a(f_i)=-f_it_i,\quad a(q^h)=q^{-h}
\quad(h\in P^*)
\]
which equip $\u$ with a Hopf algebra structure.

The generators $\{e_i,f_i,t_i=q^{h_i}|i=0,1\}$ generate a subalgebra $\Up$
of $\u$.
Drinfeld \cite{Drinfeld2} introduced a new realization of $\Up$
which is convenient for our aim:
let $A$ be an algebra generated by
$x_m^{\pm}(m\in\Z)$, $a_m(m\in\Z_{\not=0})$,
$\gamma$ and $K$ (which we refer to as Drinfeld generators)
with relations
\begin{eqnarray}
&& \gamma\hbox{: central} \nonumber \\
&& [a_m,a_n]=\delta_{m+n,0}{[2m]\ov m}{\gamma^m-\gamma^{-m}\ov q-q^{-1}},
\nonumber \\
&& [a_m,K]=0, \nonumber \\
&& Kx_m^\pm K^{-1}=q^{\pm 2}x_m^\pm, \nonumber \\
&& [a_m,x_n^\pm]=\pm{[2m]\ov m}\gamma^{\mp{|m|\ov 2}}x_{m+n}^\pm, \nonumber \\
&& x_{m+1}^\pm x_n^\pm-q^{\pm 2}x_n^\pm x_{m+1}^\pm
  =q^{\pm 2}x_m^\pm x_{n+1}^\pm-x_{n+1}^\pm x_m^\pm, \nonumber \\
&& [x_m^+,x_n^-]={1\ov q-q^{-1}}(\gamma^{{1\ov 2}(m-n)}\psi_{m+n}-
\gamma^{-{1\ov 2}(m-n)}\varphi_{m+n}),
\label{Drinfeld-relation}
\end{eqnarray}
where
\begin{eqnarray*}
\sum_{m=0}^\infty\psi_mz^{-m}
&=&K \exp\big((q-q^{-1})\sum_{m=1}^\infty a_mz^{-m}\big), \\
\sum_{m=0}^\infty\varphi_{-m}z^{m}
&=&K^{-1} \exp\big(-(q-q^{-1})\sum_{m=1}^\infty a_{-m}z^{m}\big)
\end{eqnarray*}
and $\psi_{-m}=\varphi_m=0$ for $m>0$;
here the bracket $[x,y]$ means $xy-yx$.
The equations for the formal Laurent series are
to be interpreted as a set of equations for coefficients.
The essential theorem is that the algebra $A$ is isomorphic to $\Up$;
the Chevalley generators are given by the identification
\begin{equation}
t_0=\gamma K^{-1},\quad t_1=K,\quad e_1=x_0^+,\quad f_1=x_0^-,\quad
e_0=x_1^-K^{-1},\quad f_0=Kx_{-1}^+.
\label{CD-rel}
\end{equation}
We identify the algebra $A$ with $\Up$.
The coproduct of the Drinfeld generators is known partially:
for $n\ge 0$ and $l>0$ we have \cite{CP}
\begin{eqnarray}
&& \Delta(x_n^+)=x_n^+\otimes\gamma^n+\gamma^{2n}K\otimes x_n^+
+\sum_{i=1}^{n-1}\gamma^{(n+3i)/2}\psi_{n-i}\otimes\gamma^{n-i}x_i^+
\nonumber \\
&&\qquad\qquad\qquad\qquad
\qquad\qquad\qquad\qquad\qquad\qquad\bmod N_-\otimes N_+^2, \nonumber  \\
&& \Delta(x_{-l}^+)=x_{-l}^+\otimes\gamma^{-l}+K^{-1}\otimes x_{-l}^+
+\sum_{i=1}^{l-1}\gamma^{(l-i)/2}\varphi_{-l+i}\otimes\gamma^{-l+i}x_{-i}^+
\nonumber \\
&&\qquad\qquad\qquad\qquad
\qquad\qquad\qquad\qquad\qquad\qquad\bmod N_-\otimes N_+^2, \nonumber  \\
&& \Delta(x_{l}^-)=\gamma^{l}\otimes x_{l}^-+x_{l}^-\otimes K
+\sum_{i=1}^{l-1}\gamma^{i}x_{l-i}^-\otimes\gamma^{-i/2}\psi_{i}
\nonumber \\
&&\qquad\qquad\qquad\qquad
\qquad\qquad\qquad\qquad\qquad\qquad\bmod N_-^2\otimes N_+, \nonumber  \\
&& \Delta(x_{-n}^-)=\gamma^{-n}\otimes x_{-n}^-+x_{-n}^-\otimes
\gamma^{-2n}K^{-1}
+\sum_{i=1}^{n}\gamma^{-i}x_{-n+i}^-\otimes\gamma^{-(4n-3i)/2}\varphi_{-i}
\nonumber \\
&&\qquad\qquad\qquad\qquad
\qquad\qquad\qquad\qquad\qquad\qquad\bmod N_-^2\otimes N_+, \nonumber  \\
&& \Delta(a_l)=a_l\otimes\gamma^{l/2}+\gamma^{3l/2}\otimes a_l
\bmod N_-\otimes N_+, \nonumber  \\
&& \Delta(a_{-l})=a_{-l}\otimes\gamma^{-3l/2}+\gamma^{-l/2}\otimes a_{-l}
\bmod N_-\otimes N_+; \label{D-coproduct}
\end{eqnarray}
here $N_\pm$ and $N_\pm^2$ are
left $F[\gamma^\pm,\psi_r,\varphi_s|r,-s\in\Z_{\ge 0}]$-modules generated by
$\{x_m^\pm|m\in\Z\}$ and $\{x_m^\pm x_n^\pm|m,n\in\Z\}$ respectively.
It gives sufficient information for our study
(we do not use $\Delta(x_{m}^-), m\in\Z$, in this paper).

We will use the coproduct and the antipode
to define a tensor product and dual representations, respectively.
Let $(\pi_V,V)$ be a representation.
The tensor product representation $(\pi_{V\otimes W},V\otimes W)$ is defined
to be $\pi_{V\otimes W}=(\pi_V\otimes\pi_W)\circ\Delta$.
The dual representations $(\pi_V^{*a^{\pm 1}},V^*)$ are defined to be
$\pi_V^{*a^{\pm 1}}={}^t\pi_V\circ a^{\pm 1}$;
we also refer to them as modules $V^{*a^{\pm 1}}$.
Note that the $V^{*a^{\pm 1}}$ are left modules.

\subsection{Finite-dimensional representations}
For a positive integer $k$
let $V^{(k)}=\bigoplus_{j=0}^k Fu_j$ be a $(k+1)$-dimensional vector space.
The following gives the irreducible $(k+1)$-dimensional representation
$(\pi,V^{(k)})$ of the algebra $\Up$:
\[
\pi(t_1)u_j=q^{k-2j}u_j,\quad
\pi(e_1)u_j=[j]u_{j-1},\quad
\pi(f_1)u_j=[k-j]u_{j+1},
\]
\[
\pi(t_0)=\pi(t_1)^{-1},\quad
\pi(e_0)=\pi(f_1),\quad
\pi(f_0)=\pi(e_1)
\]
or
\[
\pi(\gamma)u_j=u_j,\quad
\pi(K)u_j=q^{k-2j}u_j,\quad
\pi(a_m)u_j=a_m^ju_j,
\]
\[
\pi(x_m^+)u_j=q^{m(k-2j)}[j]u_{j-1},\quad
\pi(x_m^-)u_j=q^{m(k-2j)}[k-j]u_{j+1}
\]
where
\[
a_m^j=-{[2m][jm]\ov m[m]}q^{m(k-j+1)}+{[km]\ov m}.
\]

Put $V_z^{(k)}=V^{(k)}\otimes F[z,z^{-1}]$.
Then, the following gives a representation $(\pi_z,V^{(k)}_z)$ of $\u$:
\[
\pi_z(x)=\pi(x)\otimes{\rm id} \mbox{ for $x=e_1,f_1,t_1,t_0$},
\]
\[
\pi_z(e_0)=\pi(f_1)\otimes z,\quad
\pi_z(f_0)=\pi(e_1)\otimes z^{-1},
\]
\[
\pi_z(q^d)(u_j\otimes z^n)=q^nu_j\otimes z^n
\]
or
\[
\pi_z(\gamma)=\pi(\gamma)\otimes{\rm id},\quad
\pi_z(K)=\pi(K)\otimes{\rm id},\quad
\pi_z(a_m)=\pi(a_m)\otimes z^m,
\]
\[
\pi_z(x_m^\pm)=\pi(x_m^\pm)\otimes z^m.
\]

Later, when considering an application to a spin chain or a vertex model,
we will need the following isomorphisms between left $\u$-modules:
\begin{eqnarray*}
V^{(k)}_{zq^{-2}} &{\buildrel\sim\over\longrightarrow}& V^{(k)*a^{\pm 1}}_z \\
u_j & \longmapsto & c^\pm_ju_{k-j}^*\quad(j=0,1,\ldots,k)
\end{eqnarray*}
where
\begin{equation}
c^\pm_j=(-1)^jq^{j^2-(k\pm 1)j}
{1\ov \left[ \begin{array}{c}k\\ j\end{array} \right]}.
\label{cc}
\end{equation}
We shall denote $c_j=c^+_j$
(we do not use $c^-_j$ in this paper).

\subsection{Irreducible highest weight modules}
Set $P_+=\Z_{\ge 0}\Lambda_0+\Z_{\ge 0}\Lambda_1$.
For $\lambda\in P_+$, a $\u$-module $V(\lambda)$ is called
an irreducible highest weight module with highest weight $\lambda$
if the following conditions are satisfied:
there is a vector $\ket{\lambda}\in V(\lambda)$,
called the highest weight vector,
such that
$q^h\ket{\lambda} = q^{\br{\lambda,h}}\ket{\lambda} (h\in P^*)$,
$e_i\ket{\lambda} = f_i^{\br{\lambda,h_i}+1}\ket{\lambda}=0 (i=0,1)$,
and $V(\lambda) = \u\ket{\lambda}$.
We say that $V(\lambda)$ has level $k$ if
$\gamma\ket{\lambda}=q^k\ket{\lambda}$.
The $V(\lambda)$ has a weight-space decomposition
$V(\lambda)=\oplus_{\mu\in P}V(\lambda)_\mu$.
We write the weight of a weight vector $v$ as ${\rm wt}(v)$.
We use the notation      
$\lambda_m =(k-m)\Lambda_0+m\Lambda_1$,
$m=0,1,\ldots,k$, for level $k$ highest weights.

\subsection{Vertex operators}
Vertex operators of type~I are mappings
\begin{equation}
\Phi_{\lambda_m}^{\lambda_{k-m}V}(z):
V(\lambda_m)\longrightarrow V(\lambda_{k-m})\otimes V^{(k)}_z
\label{VO-mapping}
\end{equation}
that commute with the action of the algebra $\u$ (intertwiners), i.e.,
\begin{equation}
\Delta(x)\circ\Phi(z) = \Phi(z)\circ x \quad{\rm for}\; x\in \u.
\label{VO-intertwin}
\end{equation}
A convenient normalization is
\begin{equation}
\widetilde{\Phi}_{\lambda_m}^{\lambda_{k-m}V}(z)(\ket{\lambda_m})=\ket{\lambda_{k-m}}\otimes u_{k-m} + \ldots.
\label{VO-normalization}
\end{equation}
Let us write
\[
\Phi(z)=\sum_{j=0}^k\Phi_j(z)\otimes u_j
\]
and call $\Phi_j(z)$ the $j$-th component of $\Phi(z)$.

It was proved that vertex operators $\Phi_{\lambda_m}^{\lambda_{k-m}V}$
($m=0,1,\ldots,k$) which we defined above exist uniquely
up to normalization \cite{DJO}.

Precisely speaking, the right hand side of (\ref{VO-mapping})
is to be interpreted as
$\bigoplus_\mu\prod_\nu V(\lambda_{k-m})_\nu\otimes (V^{(k)}_z)_{\mu-\nu}$;
but we will not go into such details on topology.

\subsection{Vertex-operator $n$-point functions}
The aim of this paper is
to get an integral formula for the $n$-point function of vertex operators
\begin{equation}
\widetilde F^{(\lambda)}_{i_1,\ldots,i_n}(z_1,\ldots,z_n|\xi,y)\equiv
\tr_{V(\lambda)}(\xi^{-2d}y^{\alpha}
\widetilde\Phi_{i_1}(z_1)\cdots\widetilde\Phi_{i_n}(z_n))
\label{VO-corr}
\end{equation}
where $y^\alpha$ acts as $y^\alpha v=y^{(\alpha,{\rm wt}(v))}v$
on $v\in V(\lambda)$;
$\Phi_{i_1}(z_1)\cdots\Phi_{i_n}(z_n)$ being the $(i_1,\ldots.i_n)$-component
of a composite mapping
\begin{eqnarray*}
&&(\Phi(z_1)\otimes{\rm id}_{V\otimes\cdots\otimes V})\circ
\cdots\circ(\Phi(z_{n-1})\otimes{\rm id}_V)\circ\Phi(z_n):   \\
&&\qquad\qquad\qquad\qquad\qquad
V(\lambda)\longrightarrow
V(\lambda)\otimes V^{(k)}_{z_1}\otimes\cdots\otimes V^{(k)}_{z_n};
\end{eqnarray*}
it is understood that
$\widetilde\Phi_{i_n}(z_n): V(\lambda)\to V(\sigma(\lambda))$,
$\widetilde\Phi_{i_{n-1}}(z_{n-1}): V(\sigma(\lambda))\to V(\lambda)$, etc.,
where $\sigma(l\Lambda_0+m\Lambda_1) \equiv m\Lambda_0+l\Lambda_1$;
note that
$n$ must be even unless $k=$even and $\lambda = (k/2)(\Lambda_0+\Lambda_1)$.
$\xi$ and $y$ are free parameters.
We assume $|\xi|<1$ to make the trace convergent.

\subsection{Spin correlation functions}
We first introduce the {\it inverse} of the vertex operator.
Then
we give a formula for
a spin correlation function of an $S=k/2$ quantum spin chain
or an integrable vertex model
(which is defined to be
the vacuum expectation value of an $n$-point local operator in the spin chain,
or the thermal average of an in-row product of $n$ variables
in the vertex model).
As we shall see
{\it the spin $n$-point correlation function is obtained from
the vertex operator $2n$-point functions (\ref{VO-corr})}
by specializing parameters.

\subsubsection{Inverse vertex operators}
Let $\widetilde\Phi^{\lambda_m}_{\lambda_{k-m}V}(z)$ be vertex operators
\[
\widetilde\Phi^{\lambda_m}_{\lambda_{k-m}V}(z):
V(\lambda_{k-m})\otimes V^{(k)}_z \longrightarrow V(\lambda_m)
\]
which commute with the action of $\u$ (intertwiners)
with normalization
\[
\widetilde\Phi^{\lambda_m}_{\lambda_{k-m}V}(z)(\ket{\lambda_{k-m}}\otimes
u_{k-m})
=\ket{\lambda_m}+\ldots.
\]
(the omitted terms $\ldots$ have weight not equal to $\lambda_{m}$).
A natural identification of $V(\mu)\otimes V^{(k)}_z\to V(\lambda)$
with $V(\mu)\to V(\lambda)\otimes V^{(k)*a}_z$ (as $\u$-intertwiners)
and the isomorphism $V^{(k)}_{zq^{-2}}\simeq V^{(k)*a}_z$
yield
\begin{equation}
\widetilde\Phi^{\lambda_m}_{\lambda_{k-m}V,j}(z)
={c_{k-j}\ov c_{m}}
\widetilde\Phi^{\lambda_mV}_{\lambda_{k-m},k-j}(zq^{-2}),\quad
m=0,1,\ldots,k;
\label{inverseVO}
\end{equation}
$c_m$ being given by (\ref{cc}).
They are regarded as the inverses (up to constant factors)
of the previous ones
in the following sense (for the proof see \cite{IIJMNT}):
\begin{eqnarray*}
\widetilde\Phi^{\lambda_m}_{\lambda_{k-m}V}(z)\circ
\widetilde\Phi^{\lambda_{k-m}V}_{\lambda_m}(z)
&=& g_{\lambda_m}\times{\rm id}_{V(\lambda_m)}, \\
\widetilde\Phi^{\lambda_{k-m}V}_{\lambda_m}(z)\circ
\widetilde\Phi^{\lambda_m}_{\lambda_{k-m}V}(z)
&=& g_{\lambda_m}\times{\rm id}_{V(\lambda_{k-m})\otimes V} \\
\end{eqnarray*}
where
\begin{equation}
g_{\lambda_m}=q^{(k-m)m}\left[ \begin{array}{c}k\\ m\end{array} \right]
{\p{q^{2(k+1)};q^4}\ov \p{q^2;q^4}}.
\label{g}
\end{equation}
We shall frequently use the notation
\[
\p{z;x}=\prod_{n=1}^\infty(1-zx^{n-1}).
\]

\subsubsection{Spin correlations}
\label{sec-spin-corr-general}
The Hamiltonian of the integrable extension to spin~$k/2$
of the XXZ model (we call the spin~$k/2$ analog of the XXZ model)
is defined by
\[
H=\sum_{l\in\Z}
\cdots \otimes \head{1}^{l+2}\otimes \head{h}^{l+1\quad l} \otimes
\head{1}^{l-1}\otimes \cdots,
\]
\begin{equation}
h=(-1)^k(q^k-q^{-k}) \biggl[{d\over dz}\check R(z,1)\biggr]_{z=1};
\nonumber
\end{equation}
here $\check R$ is a $\Up$-homomorphism, called the $R$-matrix,
\[
\check R(z_1/z_2): V_{z_1}^{(k)}\otimes V_{z_2}^{(k)} \longrightarrow
V_{z_2}^{(k)}\otimes V_{z_1}^{(k)}.
\]
The $R$-matrix itself defines an integrable $k+1$-state vertex model.

In the formulation of Refs.\cite{DFJMN,JMMN,IIJMNT}
the vacuum expectation value of a local operator
$L \in {\rm End}(V^{(k)\otimes n})$
(where $V^{(k)\otimes n}$ is understood as $n$th to 1st components in
$V^{(k)\otimes\infty}$)
is given by \cite{IIJMNT}
\begin{eqnarray}
&&\br{L}^{(\lambda)}_{z_n,\ldots,z_1}
={\tr_{V(\lambda)}(q^{-2\rho}\varrho^{(\lambda)}_{z_n,\ldots,z_1}(L))
\ov \tr_{V(\lambda)}(q^{-2\rho})}
\label{local}
\end{eqnarray}
where $\rho = \La_0 + \La_1$ and
\[
\varrho^{(\lambda)}_{z_n,\ldots,z_1}(L)
=(\Phi^{(n)}_{\lambda}(z_n,\ldots,z_1)^{-1}\circ (\id_{V(\lambda^{(n)})}\otimes
L)
\circ \Phi^{(n)}_{\lambda}(z_n,\ldots,z_1)),
\]
\begin{eqnarray*}
&&\Phi^{(n)}_{\lambda}(z_n,\ldots,z_1)  \\
&&\qquad
=(\widetilde\Phi_{\lambda^{(n-1)}}^{\lambda^{(n)} V}(z_n)\otimes
\id_{\underbrace{\scriptstyle V \otimes\cdots\otimes V}_{n-1}})\circ\cdots\circ
(\widetilde\Phi_{\lambda^{(1)}}^{\lambda^{(2)} V}(z_2)\otimes\id_{V})\circ
\widetilde\Phi_{\lambda}^{\lambda^{(1)} V}(z_1),   \\
&&\Phi^{(n)}_{\lambda}(z_n,\ldots,z_1)^{-1}  \\
&&\qquad
=\widetilde\Phi^{\lambda}_{\lambda^{(1)} V}(z_1)\circ
(\widetilde\Phi^{\lambda^{(1)}}_{\lambda^{(2)} V}(z_2)\otimes\id_{V})
\circ\cdots\circ
(\widetilde\Phi^{\lambda^{(n-1)}}_{\lambda^{(n)} V}(z_n)\otimes
\id_{\underbrace{\scriptstyle V \otimes\cdots\otimes V}_{n-1}})  \\
&&\qquad\qquad
\times(g_{\lambda}g_{\lambda^{(1)}}\cdots g_{\lambda^{(n-1)}})^{-1}.
\end{eqnarray*}
A general local operator is
expressed as a linear combination of
$E_{i_nj_n}\otimes\cdots\otimes E_{i_1j_1}$;
so it is sufficient to give a formula for this basis operator
(below we shall set $\lambda = \lambda_m$):
\begin{eqnarray*}
&&\br{E_{i_nj_n}\otimes\cdots\otimes E_{i_1j_1}}^{(\lambda_m)}_{z_n,\ldots,z_1}
\nonumber\\
&&\quad=
{\tr_{V(\lambda_m)}(q^{-2\rho}
\widetilde\Phi^{\lambda_m}_{\lambda^{(1)}V,i_1}(z_1)
\cdots
\widetilde\Phi^{\lambda^{(n-1)}}_{\lambda^{(n)}V,i_n}(z_n)
\widetilde\Phi^{\lambda^{(n)}V}_{\lambda^{(n-1)},j_n}(z_n)
\cdots
\widetilde\Phi^{\lambda^{(1)}V}_{\lambda_m,j_1}(z_1))
\ov
g_{\lambda_m} g_{\lambda^{(1)}}\cdots g_{\lambda^{(n-1)}}
\tr_{V(\lambda_m)}(q^{-2\rho})};
\end{eqnarray*}
$(E_{ij})_{i'j'}=\delta_{ii'}\delta_{jj'}$; $\lambda^{(l)} =
\sigma(\lambda^{(l-1)})$, $\lambda^{(0)} = \lambda_m$.
Substituting (\ref{inverseVO})
we see that the spin $n$-point correlation function is
obtained from a vertex operator $2n$-point function by specializing
the parameters $\xi=q^2$, $y=q^{-1}$ (cf. $\rho=2d+{1\over2}\alpha$):
\begin{eqnarray}
&&\br{E_{i_nj_n}\otimes\cdots\otimes E_{i_1j_1}}^{(\lambda_m)}_{z_n,\ldots,z_1}
=(-1)^{mn+k(\floor{{n\over2}}+n)+\sum_{l=1}^{n}i_l}
\nonumber\\
&&\quad\times
q^{-m\ceil{{n\over2}}-(k-m)\floor{{n\over2}}+\sum_{l=1}^{n}(k-i_l)(1-i_l)}
{\p{q^2;q^4}^n\over\p{q^{2(k+1)};q^4}^n}
\prod_{l=1}^{n}\left[\matrix{k\cr i_l\cr}\right]^{-1}
\nonumber\\
&&\quad\times
{\widetilde F^{(\lambda_m)}_{k-i_1,\ldots,k-i_n,j_n,\ldots,j_1}
(z_1q^{-2},\ldots,z_nq^{-2},z_n,\ldots,z_1|q^2,q^{-1})
\over
\tr_{V(\lambda_m)}(q^{-2\rho})};
\label{spin-corr}
\end{eqnarray}
we have used (\ref{cc}), (\ref{g}) and introduced
$\ceil x \equiv{\rm min}\{n\in\Z|n\ge x\}$,
$\floor x \equiv{\rm max}\{n\in\Z|n\le x\}$.

We note that $z_j$ are regarded as the (trigonometric) spectral parameters
in the lattice model language.
For spin~$k/2$ analog of the XXZ model
we specialize the spectral parameters to $z_1 = \cdots = z_n$.
For the vertex model we are not necessarily to specialize them;
in this case $z_j$ is the spectral parameter of the $j$th vertical line
and the correlator is for the inhomogeneous $k+1$-state vertex model.

Finally we must mention the range of the parameter $q$.
In the application to spin correlation functions
we must assume that $|q|<1$ in order to make traces convergent.
Furthermore, we assume $-1<q<0$;
this assumption seems necessary to verify our identification of the vector
$\ket{vac}=\sum v_i\otimes v_i^* \in V(\lambda)\otimes V(\lambda)^{*a}$
with a physical vacuum (ground state)
via a discussion using the crystal base (cf. \cite{DFJMN,IIJMNT}).
We shall see, however, the final formula still has a meaning even at
positive $q$; i.e., for $|q|<1$.

\section{Constructions of level two irreducible highest weight modules over
$U_q(\widehat{sl}_2)$}
\label{sec-modules}
We shall give explicit constructions of three level two irreducible highest
weight modules,
$V(\lambda)$, $\lambda=2\Lambda_0,2\Lambda_1,\Lambda_0+\Lambda_1$.
We use the Drinfeld realization of $\Up$
in which the operators $a_m$ generate a boson algebra ${\cal A}$.
To construct boson vacuum spaces we have to introduce a fermion:
a Neveu-Schwarz fermion for $V(2\Lambda_i)$, $i=0,1$,
and a Ramond fermion for $V(\Lambda_0+\Lambda_1)$;
just as in the constructions of level two modules over the affine Lie algebra
$\widehat{sl}_2$.

We first define a total Fock space $\F$ in which the irreducible modules are
to be embedded.
The structure of the total Fock space is as follows:
$\F={\cal A}\Omega$, where $\Omega$ is a boson vacuum space
(i.e., each vector in $\Omega$ generates a boson Fock space);
the $\Omega$ is realized by a product of the group algebra of
the weight lattice of $sl_2$ and a fermion Fock space.
The aim is to single out the highest weight vector $\ket{\lambda}$ in $\F$
and to find explicit forms of the currents
$x^\pm(z)\equiv\sum_{m\in\Z}x^\pm_mz^{-m}$ as operators on $\F$
so that the $\ket{\lambda}$ and the $x^\pm(z)$ generate the
irreducible module $V(\lambda)$.

It should be noted that
the constructions were essentially obtained by Bernard
who gave in \cite{Bernard} the level one modules over $U_q(B_r^{(1)})$,
the case $r=1$ corresponding to ours;
but it seems that
an explicit exposition in the $\u$ case is not found in literatures;
so we decided  to describe it in details without proof.

{}From here we fix the level $k=2$.

\subsection{Fock spaces}
We first prepare necessary operators and Fock spaces.

Drinfeld generators $a_m, \gamma$ form a Heisenberg subalgebra ${\cal A}$
of the algebra $\Up$.
Let us put $\gamma=q^2$ since we want to construct level two modules;
then
\[
[a_m,a_n]=\delta_{m+n,0}{[2m]^2\over m},\quad m,n\in\Z_{\not=0}.
\]
We refer to the generators $a_m$ as boson;
$a_{-m}$, $m>0$, are creation operators
and $a_{m}$, $m>0$, annihilation operators.
A Fock space of the boson is defined as usual
which we denote $\F^a=F[a_{-1},a_{-2},\ldots]$:
\begin{eqnarray*}
\F^a&\equiv&F[a_{-1},a_{-2},\ldots]
=\bigoplus_{\scriptstyle
            \begin{array}{c}
              1\le i_1<\cdots<i_s \\
              n_1,\ldots,n_s>0
            \end{array}}
\; F \;a_{-i_1}^{n_1}\cdots a_{-i_s}^{n_s}\ket{},
\end{eqnarray*}
$\ket{}$ (which will be denoted $1$ in the following)
being a vacuum vector such that $a_m\ket{}=0$ for $m>0$,
with the natural left action of the boson algebra.

Let $P=\Z{\alpha\over 2}$ and $Q=\Z\alpha$ be
the weight lattice and the root lattice of
the Lie algebra $sl_2$, respectively;
the $\alpha$ being the simple root of $sl_2$, identified with $\alpha_1$.
Let $F[P],F[Q]$ be their group algebras.
Elements $e^{n{\alpha\over 2}}, n\in\Z$ span the vector space $F[P]$.
$F[Q]$ is a subspace of $F[P]$, and $F[P]=F[Q]\oplus e^{{\alpha\over 2}}F[Q]$.
We introduce two linear operators on $F[P]$ which are
$e^{\beta}(\beta\in\Z\alpha)$ and $\partial_\alpha$ defined by
\[
e^{\beta_1}\cdot
e^{\beta_2}=e^{\beta_1+\beta_2}\;(\beta_1\in\Z\alpha,\beta_2\in P),
\quad
\partial_\alpha\cdot e^\beta=(\alpha,\beta)e^\beta\;(\beta\in P).
\]

Let us introduce
the Neveu-Schwarz fermion $\{\phi_n^{NS}|n\in\Z+{1\ov 2}\}$,
and the Ramond fermion $\{\phi_n^{R}|n\in\Z\}$;
both satisfy the anti-commutation relation
\[
[\phi_m,\phi_n]_+=\delta_{m+n,0}{q^{2m}+q^{-2m}\over q+q^{-1}}.
\]
We have dropped the indices $NS$, $R$.
Elements $\phi_{-n}$, $n>0$, are creation operators
and $\phi_{n}$, $n>0$, annihilation operators.
Construct Fock spaces for these fermions as usual and
denote them
$\F^{\phi^{NS}}=F[\phi_{-1/2},\phi_{-3/2},\ldots]$,
$\F^{\phi^{R}}=F[\phi_{-1},\phi_{-2},\ldots]$.

One should notice that $\phi_0$ of the Ramond is special
since it anti-commutes with
$\phi_n$ ($n\in\Z_{\not= 0}$) and $\phi_0^2=1/[2]$;
we must treat the Ramond fermion carefully.
It will be convenient to write the Ramond fermion as
\[
\phi_n=\psi_n\otimes\left(\matrix{1&0\cr0&-1\cr}\right)\;(n\not= 0),\quad
\phi_0=\psi_0\otimes\left(\matrix{0&1\cr1&0\cr}\right);
\]
\[
[\psi_m,\psi_n]_+=\delta_{m+n,0}{q^{2m}+q^{-2m}\over q+q^{-1}}\;
(m,n\in\Z_{\not=0}),\quad
\psi_0=[2]^{-{1\over 2}}\in F.
\]
Then
\[
\F^{\phi^R}=\F^\psi\otimes\C^2=F[\psi_{-1},\psi_{-2},\ldots]\otimes\C^2.
\]
A convenient basis of $\C^2$ is
$\{\left(\matrix{1\cr\epsilon\cr}\right)|\epsilon=\pm1\}$;
the vectors are eigenvectors of $\phi_0$ and flipped to each other by
$\phi_n(n\not=0)$ since
\[
\left(\matrix{0&1\cr1&0\cr}\right)\left(\matrix{1\cr\epsilon\cr}\right)=
\epsilon\left(\matrix{1\cr\epsilon\cr}\right),\quad
\left(\matrix{1&0\cr0&-1\cr}\right)\left(\matrix{1\cr\epsilon\cr}\right)=
\left(\matrix{1\cr-\epsilon\cr}\right).
\]

Define two vector spaces by
\begin{eqnarray}
\F^{(0)}&\equiv&\F^a\otimes\F^{\phi^{NS}}\otimes F[Q],
\label{total-Fock-0}\\
\F^{(1)}&\equiv&\F^a\otimes\F^{\phi^{R}}\otimes e^{{\alpha\over 2}}F[Q]
\label{total-Fock-1}
\end{eqnarray}
which we shall refer to as total Fock spaces.
The boson, fermion, and lattice operators naturally act on the spaces
(the boson acts on the first component, etc.).
As we shall see  modules $V(2\Lambda_0)$ and $V(2\Lambda_1)$
are contained in $\F^{(0)}$ and
a module $V(\Lambda_0+\Lambda_1)$ is contained in $\F^{(1)}$.

\subsection{Irreducible highest weight modules $V(2\Lambda_i)$, $i=0,1$}
We first consider an irreducible highest weight module with highest weight
$2\Lambda_0$ or $2\Lambda_1$.

We define the action of Drinfeld generators $\gamma$ and $K$ as
\begin{equation}
\gamma=q^2,\quad K=q^{\partial_\alpha}.
\end{equation}
Then, from the Drinfeld relations (\ref{Drinfeld-relation}),
explicit forms of currents (or generating functions)
\[
x^\pm(z)=\sum_{m\in\Z}x_m^\pm z^{-m}
\]
as operators on the total Fock space $\F^{(0)}$
are determined to be
\begin{eqnarray}
&&x^\pm(z)=[2]^{{1\over 2}}
E^\pm_<(z)E^\pm_>(z)
\phi(z)e^{\pm\alpha}z^{{1\over 2}\pm{1\over 2}\partial_\alpha};
\label{current}
\end{eqnarray}
\begin{eqnarray*}
&&E^\pm_<(z)=\exp(\pm\sum_{m=1}^\infty{a_{-m}\over[2m]}q^{\mp m}z^{m}),\quad
E^\pm_>(z)=\exp(\mp\sum_{m=1}^\infty{a_{m}\over[2m]}q^{\mp m}z^{-m});
\end{eqnarray*}
$\phi(z)$ being the Neveu-Schwarz fermion field
\[
\phi(z)=\sum_{n\in\Z+{1\over 2}}\phi_n z^{-n};
\]
equations (\ref{current}) are to be understood as a set of equations
for coefficients.
We have defined the action of the algebra $\Up$ on the total Fock space
$\F^{(0)}$;
i.e., $\F^{(0)}$ is a $\Up$-module.

A vector $1$ with weight $2\Lambda_0$
(cf. $\gamma\cdot 1=q^2 1$, $K\cdot 1=1$)
generates a $\Up$-submodule
\begin{eqnarray*}
\F^{(0)}_+&=&
\F^a\otimes
\Big\{
\big(\F^{\phi^{NS}}_{even}\otimes F[2Q]\big)\oplus
\big(\F^{\phi^{NS}}_{odd}\otimes e^\alpha F[2Q]\big)
\Big\};
\end{eqnarray*}
this subspace of $\F^{(0)}$ is identified with $V(2\Lambda_0)$,
the highest weight vector being $1$.

A vector $e^\alpha$ with weight $2\Lambda_1$
(cf. $\gamma\cdot e^\alpha=q^2 e^\alpha$, $K\cdot e^\alpha=q^2e^\alpha$)
generates a $\Up$-submodule
\begin{eqnarray*}
\F^{(0)}_-&=&
\F^a\otimes
\Big\{
\big(\F^{\phi^{NS}}_{even}\otimes q^\alpha F[2Q]\big)\oplus
\big(\F^{\phi^{NS}}_{odd}\otimes F[2Q]\big)
\Big\};
\end{eqnarray*}
this subspace is identified with $V(2\Lambda_1)$,
the highest weight vector being $e^\alpha$.

Note that the total Fock space is a direct sum of the two subspaces
\[
\F^{(0)}=\F^{(0)}_+\oplus\F^{(0)}_-.
\]

Define the operator $d$ by
\begin{eqnarray}
&&d=-\sum_{m=1}^\infty mN_m^a-\sum_{k>0}kN_k^\phi
-{1\over 8}\partial_\alpha^2+{(\lambda,\lambda)\over 4}
\label{d}
\end{eqnarray}
where
\begin{equation}
N_m^a={m\over[2m]^2}a_{-m}a_{m},\quad
N_m^\phi={q+q^{-1}\over q^{2m}+q^{-2m}}\phi_{-m}\phi_{m} \quad(m>0).
\label{N}
\end{equation}
This operator is identified with the grading operator $d$ in the algebra $\u$;
with this $d$ and setting $\lambda=2\Lambda_i$
we have irreducible highest weight modules $V(2\Lambda_i)\simeq\F^{(0)}_\pm$
($\pm$ according to $i=0,1$) over $\u$.

\subsection{Irreducible highest weight module $V(\Lambda_0+\Lambda_1)$}
Now we consider an irreducible highest weight module with highest weight
$\Lambda_0+\Lambda_1$.

The action of Drinfeld generators $\gamma$ and $K$ is defined as before.
Explicit forms of currents
as operators on the total Fock space $\F^{(1)}$
are the same as (\ref{current}) except for the fermion field
to be replaced by the Ramond one
\[
\phi(z)=\sum_{n\in\Z}\phi_n z^{-n}.
\]
Thus we have got a $\Up$-module $\F^{(1)}$.

Fix a one of the vectors
$\left(\matrix{1\cr\epsilon\cr}\right)$, $\epsilon=\pm1$.
A vector $\left(\matrix{1\cr\epsilon\cr}\right)\otimes e^{{\alpha\over 2}}$
with weight $\Lambda_0+\Lambda_1$
(cf. values of $\gamma$ and $K$ on this vector are $q^2$ and $q$ respectively)
generates a $\Up$-submodule
\begin{equation}
\F^{(1)}_\epsilon=
\F^a\otimes
\Big\{
\big(\F^{\psi^{R}}_{even}\otimes \left(\begin{array}{c}1\\
\ep\end{array}\right)\big)\oplus
\big(\F^{\psi^{R}}_{odd}\otimes \left(\begin{array}{c}1\\
-\ep\end{array}\right)\big)
\Big\}\otimes e^{{\alpha\ov 2}}F[Q];
\label{total-Fock-1-ep}
\end{equation}
this subspace of $\F^{(1)}$ is identified with $V(\Lambda_0+\Lambda_1)$
with the highest weight vector
$\left(\matrix{1\cr\epsilon\cr}\right)\otimes e^{{\alpha\over 2}}$.

The total Fock space is a direct sum of the two subspaces $\F^{(1)}_\epsilon$
\[
\F^{(1)}=\F^{(1)}_+\oplus\F^{(1)}_-.
\]

Introducing the operator $d$ defined by (\ref{d}) with
$\lambda=\Lambda_0+\Lambda_1$ and the fermion being the Ramond one
we have an irreducible highest weight modules
$V(\Lambda_0+\Lambda_1)\simeq\F^{(1)}_\epsilon$ over $\u$.

\subsection{Characters}
Characters
\[
\tr_{V(\lambda)}(\xi^{-2d}y^{\partial_\alpha})
\]
for irreducible highest weight modules $V(\lambda)$,
$\lambda=2\Lambda_0,2\Lambda_1,\Lambda_0+\Lambda_1$,
can be computed directly:
\begin{eqnarray*}
&&\tr_{V(2\Lambda_i)}(\xi^{-2d}y^{\partial_\alpha})
=\xi^{-{1\over2}(2\Lambda_i,2\Lambda_i)}\times
\Big[\tr_{\F^{(0)}}(\xi^{-2d'}y^{\partial_\alpha})\Big]^{\sigma_i} \\
&&\qquad=\xi^{-{1\over2}(2\Lambda_i,2\Lambda_i)}\times
\Big[{1\over\p{\xi^2;\xi^2}}\p{-\xi;-\xi}\p{-\xi y^2;\xi^2}
\p{-\xi y^{-2};\xi^2}\Big]^{\sigma_i}, \\
&&\tr_{V(\Lambda_0+\Lambda_1)}(\xi^{-2d}y^{\partial_\alpha})
=y\p{-\xi^2;\xi^2}\p{-\xi^2y^2;\xi^2}\p{-y^{-2};\xi^2}
\end{eqnarray*}
where $d'=d-(\lambda,\lambda)/4$ in $V(\lambda)$,
and $\sigma_i$ indicates operations on arbitrary functions of $\xi$
\begin{equation}
f(\xi)^{\sigma_i}\equiv{1\over 2}(f(\xi)\pm f(-\xi))\quad
(\hbox{$\pm$ according to $i=0,1$}).
\label{symmetrization}
\end{equation}
In particular, a specialization $\xi=q^2$, $y=q^{-1}$ yields
$\tr_{V(2\Lambda_i)}(q^{-2\rho})=q^{-2i}\p{-q^2;q^2}\p{-q^4;q^4}$,
$\tr_{V(\Lambda_0+\Lambda_1)}(q^{-2\rho})=q^{-1}\p{-q^2;q^2}\p{-q^2;q^4}$.

\subsection{Boson calculus, fermion calculus}
Later,
when we prove intertwining relations  for vertex operators
and calculate $n$-point functions,
we shall need several formulae concerning calculus
on $a_m$ and $\phi_m$.
We have assembled some of them in Appendix:
formulae for operator product expansions, normal ordering, and traces.

\section{Vertex operators}
\label{sec-VO}
We give expressions of level two vertex operators
associated with our explicit constructions of level two
irreducible highest weight modules over $\u$;
see (\ref{VO-mapping}), (\ref{VO-intertwin}) with $k=2$.
First we present explicit forms of components $\Phi_j(z)$;
then we give the proof.

\subsection{Intertwining relations}
By definition the vertex operator must satisfy
intertwining relations (\ref{VO-intertwin}).
Let us write down the relations for the Chevalley generators
(we write them in terms of $x^\pm_n$; cf.(\ref{CD-rel})):
\begin{eqnarray}
f_1:\qquad
&&[\Phi_2(z),x^-_0]_{q^2}=\Phi_1(z)\label{def-Phi_1}\\
&&[\Phi_1(z),x^-_0]=[2]\Phi_0(z)\label{def-Phi_0}\\
&&[\Phi_0(z),x^-_0]_{q^{-2}}=0\\
f_0:\qquad
&&[\Phi_2(z),x^+_{-1}]=0\\
&&[\Phi_1(z),x^+_{-1}]=[2]z^{-1}t_1^{-1}\Phi_2(z)\\
&&[\Phi_0(z),x^+_{-1}]=q^{-2}z^{-1}t_1^{-1}\Phi_1(z)\\
e_1:\qquad
&&[\Phi_2(z),x^+_{0}]=0\label{check0}\\
&&[\Phi_1(z),x^+_{0}]=[2]t_1\Phi_2(z)\label{check2}\\
&&[\Phi_0(z),x^+_{0}]=t_1\Phi_1(z)\label{check3}\\
e_0:\qquad
&&[\Phi_2(z),x^-_1]_{q^{-2}}=zq^2\Phi_1(z)\\
&&[\Phi_1(z),x^-_1]=[2]zq^4\Phi_0(z)\\
&&[\Phi_0(z),x^-_1]_{q^{2}}=0\label{check1}
\end{eqnarray}
and
\begin{equation}
t_1^{-1}\Phi_j(z)t_1=q^{2-2j}\Phi_j(z),\quad
t_0^{-1}\Phi_j(z)t_0=q^{-2+2j}\Phi_j(z).
\label{intertwin-t}
\end{equation}
Here $[A,B]_x\equiv AB-xBA$.

\subsection{Explicit forms}
Components of the (unnormalized) vertex operator,
which satisfies intertwining relations, are
proved to be
\begin{eqnarray}
&&\Phi_2(z)=F_<(z)F_>(z)e^\alpha(-q^4z)^{{1\ov 2}\partial_\alpha},
\nonumber\\
&&\Phi_1(z)=\oi1D_1(z,w_1)\phi(w_1) :\Phi_2(z)\widehat{x^-}(w_1):,
\label{VO}\\
&&\Phi_0(z)=\oi2\oi1D_0(z,w_1,w_2)\phi(w_1)\phi(w_2)
:\Phi_2(z)\widehat{x^-}(w_1)\widehat{x^-}(w_2):,
\nonumber
\end{eqnarray}
where
\[
F_<(z)=\exp\big(\sum_{m=1}^\infty{a_{-m}\over[2m]}q^{5m}z^m\big),\quad
F_>(z)=\exp\big(-\sum_{m=1}^\infty{a_{m}\over[2m]}q^{-3m}z^{-m}\big),
\]
\[
D_1(z,w_1)={1-q^4\ov -q^4z}{1\ov w_1(1-q^{-2}{w_1\ov z})(1-q^6{z\ov w_1})},
\quad |{w_1\ov q^2z}|,|{q^6z\ov w_1}|<1,
\]
\[
D_0(z,w_1,w_2)={(1-q^2)^2\ov q^7z^2}{1-q^2{w_2\ov w_1}\ov
w_2\prod_{j=1,2}(1-q^{-2}{w_j\ov z})(1-q^6{z\ov w_j})},
\quad |{w_j\ov q^2z}|,|{q^6z\ov w_j}|<1;
\]
the contours for the $w$-integrals are anti-clockwise in the regions
indicated after the definitions of $D_0$ and $D_1$;
the $\hat{x}^\pm(w)$ are currents without fermion:
\begin{eqnarray*}
\hat{x}^\pm(z)
&=&[2]^{{1\ov 2}}E_<^\pm(z)E_>^\pm(z)
e^{\pm\alpha} z^{{1\over 2}\pm{1\over 2}\partial_\alpha}.
\end{eqnarray*}
The expressions for $\Phi_1,\Phi_0$ were obtained from $\Phi_2$
using (\ref{def-Phi_1}) and (\ref{def-Phi_0}).

$\Phi_j(z)$ are regarded as mappings $\F^{(i)}\to\F^{(i)}$
(we must choose the NS, R fermion for $i=0,1$, respectively)
as well as $V(\lambda)\to V(\sigma(\lambda))$;
$\sigma$ being defined by
$\sigma((2-m)\Lambda_0+m\Lambda_1)=m\Lambda_0+(2-m)\Lambda_1$, $m=0,1,2$.

The vertex operators with normalizations (\ref{VO-normalization}) are given by
\begin{eqnarray}
\widetilde{\Phi}_{2\Lambda_0}^{2\Lambda_1V}(z)&=&\Phi(z),\nonumber\\
\widetilde{\Phi}_{\Lambda_0+\Lambda_1}^{\Lambda_0+\Lambda_1V}(z)&=&\ep(-q^4z)^{{1\ov2}}\Phi(z),\nonumber\\
\widetilde{\Phi}_{2\Lambda_1}^{2\Lambda_0V}(z)&=&(-q^4z)\Phi(z);
\label{nVO}
\end{eqnarray}
at the second identification we have chosen the realization
$V(\Lambda_0+\Lambda_1)=\F^{(1)}_\ep$.
The factors were derived by applying our $\Phi(z)$ to
the highest weight vectors in our representations.

The equation for $\Phi_2$ is derived (guessed) as follows:
recall the coproducts for the Drinfeld generators (\ref{D-coproduct}) and
write down intertwining relations for $a_m$, $x_n$ and $K$;
then we have equations to be satisfied by $\Phi_2$:
\begin{eqnarray*}
&&a_m\Phi_k(z)-\Phi_k(z)a_m=q^{({3k\ov 2}+2)m}{[km]\ov m}z^m\Phi_k(z)
\quad{\rm for }\; m>0,\\
&&a_{-m}\Phi_k(z)-\Phi_k(z)a_{-m}=q^{-({k\ov 2}+2)m}{[km]\ov m}z^{-m}\Phi_k(z)
\quad{\rm for }\; m>0,\\
&&\Phi_k(z)x^+(w)-x^+(w)\Phi_k(z)=0,\\
&&K\Phi_k(z)K^{-1}=q^k\Phi_k(z);
\end{eqnarray*}
put $k=2$ (level) for our aim;
first two equations determine the boson part of $\Phi_2(z)$
and the last one requires $e^\alpha$;
the third equation is satisfied by our $\Phi_2(z)$,
so we expect that our $\Phi_2(z)$ is probably correct.

In fact the vertex operator we proposed does actually satisfy
all the intertwining relations,
which we will prove now.

\subsection{Proof of the intertwining relations for the Chevalley generators}
We prove here that our vertex operator does satisfy intertwining relations
for the Chevalley generators.
We use the operator product expansion
\[
\phi(w_1)\phi(w_2) = \br{\phi(w_1)\phi(w_2)} + :\phi(w_1)\phi(w_2):,
\quad\hbox{etc.};
\]
see Appendix A.2, A.3 for definitions and notations.

Clearly the relations (\ref{intertwin-t}) are satisfied by our $\Phi_j$.
Among the rest relations we shall prove here four typical ones which are
(\ref{check0}), (\ref{check1}), (\ref{check2}) and (\ref{check3});
one can prove other relations in almost similar ways.

\bigskip
\noindent
{\em Proof.  }
(\ref{check0}):
it holds trivially since $[\Phi_2(z),x^+(w)]=0$.

(\ref{check1}):
\begin{eqnarray*}
&&[\Phi_0(z),x^-_1]_{q^2}
=\oint{dw\ov 2\pi i}[\Phi_0(z),x^-(w)]_{q^2} \\
&&\qquad=\oi1\oi2\oi3{(1-q^2)^2\ov-q^{11}z^3} \\
&&\qquad\qquad
\times{\p{1-q^2{w_2\ov w_1}}\p{1-q^2{w_3\ov w_1}}\p{1-q^2{w_3\ov w_2}}\ov
\prod_{j=1,2,3}\{\p{1-q^{-2}{w_j\ov z}}\p{1-q^6{z\ov w_j}}\}} \\
&&\qquad\qquad
\times\Big\{w_1(1-q^2{w_2\ov w_3})+q^2{w_1w_2\ov w_3}(1-q^2{w_1\ov w_2})\Big\}
\\
&&\qquad\qquad
\times \phi(w_1)\phi(w_2)\phi(w_3)
 :\Phi_2(z)\hat x^-(w_1)\hat x^-(w_2)\hat x^-(w_3):
\end{eqnarray*}
Expand the $\{\cdots\}$ and first examine the first term;
it is an odd function with respect to $w_2\leftrightarrow w_3$
and the $w_2$-, $w_3$-integrals are independent
[note that the factor $(1-q^2{w_3\ov w_2})(1-q^2{w_2\ov w_3})$ cancels the
poles at ${w_3\ov w_2}=q^{\pm 2}$ of $\br{\phi(w_2)\phi(w_3)}$];
therefore this term vanishes after $w_2$-, $w_3$-integrations.
Similarly the second term vanishes after $w_1$-, $w_2$-integrations.
Therefore we get $[\Phi_0(z),x^-_1]_{q^2}=0$.

(\ref{check2}): we will need the operator product expansion (OPE)
of $\phi(w)\phi(w_1)$.
\begin{eqnarray*}
&&[\Phi_1(z),x^+_0]=\oint {dw\ov 2\pi i}{1\ov w}[\Phi_1(z),x^+(w)] \\
&&\qquad
=\oint_{C_1}{dw\ov2\pi i}F(w){\phi(w_1)\phi(w)\ov w_1-w}
-\oint_{C_2}{dw\ov2\pi i}F(w){\phi(w)\phi(w_1)\ov w-w_1} \\
&&\qquad
=\oint_{C_1}{dw\ov2\pi i}F(w){\br{\phi(w_1)\phi(w)}\ov w_1-w}
-\oint_{C_2}{dw\ov2\pi i}F(w){\br{\phi(w)\phi(w_1)}\ov w-w_1} \\
&&\qquad\qquad
+\oint_{C_1}{dw\ov2\pi i}F(w){:\phi(w_1)\phi(w):\ov w_1-w}
-\oint_{C_2}{dw\ov2\pi i}F(w){:\phi(w)\phi(w_1):\ov w-w_1};
\end{eqnarray*}
$C_1$ being a contour in a region $|w|<|q^{\pm 2}w_1|$,
$C_2$ in $|w|>|q^{\mp 2}w_1|$ and
\[
F(w)=\oi1{1-q^4\ov-q^4z}{(w-q^4z):\Phi_2(z)\hat x^-(w_1)\hat x^+(w):\ov
ww_1(1-q^{-2}{w_1\ov z})(1-q^6{z\ov w_1})}.
\]
The last integral can be replaced by an integral along $C_1$ plus
a residue at $w=w_1$; the former together with the third term,
and the residue, vanish respectively
because of the anti-symmetry of $:\phi(w_1)\phi(w):$.
In the same way,
deforming the contour $C_2$ of the second integral and
combining with the first term,
we have
\begin{eqnarray*}
&&[\Phi_1(z),x^+_0] \\
&&\quad
= -{\rm Res}_{w=q^2w_1}\Big\{F(w){\br{\phi(w)\phi(w_1)}\ov w-w_1} \Big\}
-{\rm Res}_{w=q^{-2}w_1}\Big\{F(w){\br{\phi(w)\phi(w_1)}\ov w-w_1} \Big\}\\
&&\quad
=\oi1{q\ov w_1(w_1-q^6z)}:\Phi_2(z)\hat x^-(w_1)\hat x^+(q^2w_1):\\
&&\quad\qquad
+\oi1{1\ov -qw_1(w_1-q^2z)}:\Phi_2(z)\hat x^-(w_1)\hat x^+(q^{-2}w_1):\\
&&\quad
=[2]q^2F_<(z)F_>(z)e^\alpha(-q^6z)^{{1\ov 2}\partial_\alpha} \\
&&\quad\qquad
\times\oint_{|q^6z|<|w_1|}{dw_1\ov2\pi i}{1\ov w_1-q^6z}
\exp(\sum_{m>0}{a_m\ov[2m]}(q^m-q^{-3m})w_1^{-m}) \\
&&\quad\qquad
-\oint_{|w_1|<|q^2z|}{1\ov w_1-q^2z}
\exp(-\sum_{m>0}{a_{-m}\ov[2m]}(q^m-q^{-3m})w_1^{m}) \times(\cdots).\\
\end{eqnarray*}
The first integral gives the residue at $\infty$ which is $1$,
and the second the residue at $w_1=0$ which is $0$;
hence
\[
[\Phi_1(z),x^+_0]=
[2]q^2F_<(z)F_>(z)q^\alpha(-q^6z)^{{1\ov 2}\partial_\alpha}
=[2]t_1\Phi_2(z);
\]
thus we have proved the relation (\ref{check2}).
Keys are
\[
:\hat x^-(w_1)\hat x^+(q^{\pm 2}w_1):\propto
\exp(\pm\sum_{m>0}{a_{\pm m}\ov[2m]}(q^m-q^{-3m})w_1^{\mp m})
\]
(does not contain boson creation, annihilation operators, respectively)
which reduce the contour integrals to simple residue calculations.

(\ref{check3}):
the proof will go along the same way as the one for (\ref{check2});
but in this case we will need the OPE of $\phi(w)\phi(w_1)\phi(w_2)$.
\begin{eqnarray*}
&&[\Phi_0(z),x^+_0]=\oint{dw\ov2\pi i}{1\ov w}[\Phi_0(z),x^+(w)] \\
&&\quad
=\oint_{C_1}{dw\ov2\pi i}F(w){\phi(w_1)\phi(w_2)\phi(w)\ov(w_1-w)(w_2-w)}
-\oint_{C_2}{dw\ov2\pi i}F(w){\phi(w)\phi(w_1)\phi(w_2)\ov(w-w_1)(w-w_2)}
\end{eqnarray*}
where
\begin{eqnarray*}
C_1:&& |w|<|w_1|,|w_2| \quad {\rm and}\quad |w|<|q^{\pm 2}w_2|; \\
C_2:&& |w|>|w_1|,|w_2| \quad {\rm and}\quad |w|>|q^{\pm 2}w_1|
\end{eqnarray*}
(the composition of fermions $\phi(w_1)\phi(w_2)\phi(w)$ requires
the second condition for $C_1$; and similarly we need the one for $C_2$)
and
\begin{eqnarray*}
F(w)&=&\oi1\oi2{(1-q^2)^2\ov q^7z^2}\\
&&\times
{(w-q^4z)(w_1-q^2w_2)
:\Phi_2(z)\hat x^-(w_1)\hat x^-(w_2)\hat x^+(w):\ov
ww_1w_2\prod_{j=1,2}\{(1-q^{-2}{w_j\ov z})(1-q^6{z\ov w_j})\}}.
\end{eqnarray*}
Substituting the OPE
\begin{eqnarray*}
\phi(w_1)\phi(w_2)\phi(w_3)
&=&\br{\phi(w_1)\phi(w_2)}\phi(w_3)-\br{\phi(w_1)\phi(w_3)}\phi(w_2) \\
&&+\br{\phi(w_2)\phi(w_3)}\phi(w_1)+:\phi(w_1)\phi(w_2)\phi(w_3):,
\end{eqnarray*}
deforming the contour $C_2$ to $C_1$ plus ones around poles,
as in the proof of (\ref{check2}),
we get
\begin{eqnarray*}
&&[\Phi_0(z),x^+_0]
=\oi1\oi2{(1-q^2)^2\ov q^7z^2}
{1\ov \prod_{j=1,2}\{(1-q^{-2}{w_j\ov z})(1-q^6{z\ov w_j})\}} \\
&&\quad
\times\Big\{
-{[2]q^4z\ov1-q^4}\phi(w_2):\Phi_2(z)\hat x^-(w_2):q^{\partial_\alpha} \\
&&\qquad\qquad\times
{(1-q^{-2}{w_1\ov z})(w_1-q^2w_2)\ov w_1w_2(w_1-q^{-2}w_2)}
\exp(\sum_{m>0}{a_{m}\ov[2m]}(q^m-q^{-3m})w_1^{-m}) \\
&&\qquad
-{[2]q^2\ov1-q^4}{(1-q^{6}{z\ov w_1})\ov w_2}
\exp(-\sum_{m>0}{a_{-m}\ov[2m]}(q^m-q^{-3m})w_1^{m}) \\
&&\qquad\qquad\times
\phi(w_2):\Phi_2(z)\hat x^-(w_2):q^{-\partial_\alpha} \\
&&\qquad
-{[2]q^6z\ov1-q^4}\phi(w_1):\Phi_2(z)\hat x^-(w_1):q^{\partial_\alpha} \\
&&\qquad\qquad\times
{(1-q^{-2}{w_2\ov z})\ov w_1w_2}
\exp(\sum_{m>0}{a_{m}\ov[2m]}(q^m-q^{-3m})w_2^{-m}) \\
&&\qquad
-{[2]\ov1-q^4}{(1-q^{6}{z\ov w_2})(w_1-q^2w_2)\ov w_1(w_1-q^{-2}w_2)}
\exp(-\sum_{m>0}{a_{-m}\ov[2m]}(q^m-q^{-3m})w_2^{m}) \\
&&\qquad\qquad\times
\phi(w_1):\Phi_2(z)\hat x^-(w_1):q^{-\partial_\alpha}\Big\} \\
\end{eqnarray*}
where contours should be $|q^6z|<|w_j|<|q^2z|$ and $|w_2/w_1|<|q^{\pm 2}|$.
The $w_1$-integral in the first term gives the residue at $w_1=\infty$;
the one in the second the residue at $w_1=0$ (which is $0$);
the $w_2$-integral in the third term gives the residue at $w_2=\infty$;
the one in the last the residue at $w_2=0$ (which is $0$).
Thus, the first and the third terms remain non-zero and give
the right hand side of the equation (\ref{check3}).

As was noted, the other relations are proved similarly.
\hfill[]
\bigskip

We have completed the proof of the intertwining relations;
hence we have proved that our expression of the vertex operator is correct.

\section{Integral formula for $n$-point functions of vertex operators}
\label{sec-formula}
In this section
we derive an integral formula for the $n$-point function of vertex operators
(VO $n$-point function)
defined in \ref{sec-formula-def}
(the trace of the product of vertex operators).
The final formula is given in
(\ref{integral-formula-0}) and (\ref{integral-formula-1}).
All the two-point functions are shown in \ref{sec-formula-example}
as an example.

\subsection{VO $n$-point functions (definition)}
\label{sec-formula-def}
Introduce more notations
\[
d'=d^a+d^\phi-{1\ov 8}\partial_\alpha^2, \quad
d^a=-\sum_{m=1}^\infty mN_m^a,\quad
d^\phi=-\sum_{m>0} mN_m^\phi;
\]
the $d'$ is related to the grading operator $d$ by
$d=d'+{1\ov 4}(\lambda,\lambda)$
in the representation $V(\lambda)$
(cf. (\ref{d}); see (\ref{N}) for definitions of $N_m^{a,\phi}$).

We shall give an integral formula for the following traces
over Fock spaces
(\ref{total-Fock-0}) and (\ref{total-Fock-1-ep}):
\begin{eqnarray}
\bF^{(0)}_{i_1,\ldots,i_n}(z_1,\ldots,z_n|\xi,y)
&\equiv&\tr_{\F^{(0)}}
(\xi^{-2d'}y^{\partial_\alpha}\Phi_{i_1}(z_1)\cdots\Phi_{i_n}(z_n)),
\nonumber\\
\bF^{(1,\ep)}_{i_1,\ldots,i_n}(z_1,\ldots,z_n|\xi,y)
&\equiv&\tr_{\F^{(1)}_\ep}
(\xi^{-2d'}y^{\partial_\alpha}\Phi_{i_1}(z_1)\cdots\Phi_{i_n}(z_n))
\label{def-bF}
\end{eqnarray}
where the vertex operators are unnormalized ones (\ref{VO}).
Since $d'$ is bounded from above in Fock spaces
we assume $|\xi|<1$ to make the traces convergent.

The $n$-point functions of the normalized vertex operators (\ref{nVO})
are defined by
\begin{eqnarray}
\widetilde{F}^{(\lambda)}_{i_1,\ldots,i_n}(z_1,\ldots,z_n|\xi,y)
&\equiv&\tr_{V(\lambda)}
(\xi^{-2d}y^{\partial_\alpha}
\widetilde{\Phi}_{i_1}(z_1)\cdots\widetilde{\Phi}_{i_n}(z_n)).
\label{def-widetildeF}
\end{eqnarray}
They are obtained from $\bF$ as:
\begin{eqnarray*}
&&\widetilde{F}^{(2\Lambda_0)}_{i_1,\ldots,i_n}(z_1,\ldots,z_n|\xi,y)\\
&&\quad=\xi^{-{1\ov 2}(2\Lambda_0,2\Lambda_0)}\cdot
(-q^4z_1)(-q^4z_3)\cdots(-q^4z_{n-1})
\Big[
\bF^{(0)}_{i_1,\ldots,i_n}(z_1,\ldots,z_n|\xi,y)
\Big]^{\sigma_0},    \\
&&\widetilde{F}^{(2\Lambda_1)}_{i_1,\ldots,i_n}(z_1,\ldots,z_n|\xi,y)\\
&&\quad=\xi^{-{1\ov 2}(2\Lambda_1,2\Lambda_1)}\cdot
(-q^4z_2)(-q^4z_4)\cdots(-q^4z_{n})
\Big[
\bF^{(0)}_{i_1,\ldots,i_n}(z_1,\ldots,z_n|\xi,y)
\Big]^{\sigma_1},    \\
&&\widetilde{F}^{(\Lambda_0+\Lambda_1)}_{i_1,\ldots,i_n}(z_1,\ldots,z_n|\xi,y)\\
&&\quad=\xi^{-{1\ov 2}(\Lambda_0+\Lambda_1,\Lambda_0+\Lambda_1)}\cdot
\epsilon^n
\prod_{l=1}^{n}(-q^4z_{l})^{{1\ov 2}}\cdot
\bF^{(1,\ep)}_{i_1,\ldots,i_n}(z_1,\ldots,z_n|\xi,y);
\end{eqnarray*}
$\sigma_i$ being defined in (\ref{symmetrization}).

If the trace is non-zero, $n$ must be even for
$\bF^{(0)}$ and $\widetilde F^{(2\Lambda_i)}$;
it is not necessarily so for $\bF^{(1,\ep)}$ and $\widetilde
F^{(\Lambda_0+\Lambda_1)}$.

\subsection{Traces}
\label{sec-formula-traces}
Before taking the trace in the $n$-point function (\ref{def-bF})
it is better to make everything in normal order.
Then
we compute the trace
separately for boson, fermion, and lattice part
(cf. Appendix for boson and fermion).
The computations are straightforward
(it is not necessary to introduce an auxiliary boson
as in Ref.\cite{JMMN}).
Here we only quote results for fermion
in order to fix the notations.

Fermion traces are taken over the Fock space
$\F^{\phi^{NS}}$ or
\begin{equation}
\F^{\phi^{R}}_\epsilon \equiv
\big(\F^{\psi^{R}}_{even}\otimes \left(\begin{array}{c}1\\
\ep\end{array}\right)\big)\oplus
\big(\F^{\psi^{R}}_{odd}\otimes \left(\begin{array}{c}1\\
-\ep\end{array}\right)\big)
\quad
(\ep = \pm 1)
\label{Ramond-Fock-space}
\end{equation}
according to Neveu-Schwarz or Ramond fermion.
A fermion trace is
expressed by a Pfaffian of a matrix $G$.
\begin{eqnarray}
&&\tr(\xi^{-2d^\phi}\phi(w_1)\cdots\phi(w_n))
\nonumber\\
&&\qquad=
\left\{\matrix{
\tr(\xi^{-2d^\phi})\times \Pf( G(\{w\}) )\hfill & \hbox{if }
  \left\{\matrix{\hbox{$n=$even,  or }\hfill\cr
                 \hbox{$n=$odd and $\phi=\phi^{R}$,} \hfill\cr}\right.\hfill\cr
0\hfill &  \hbox{otherwise.}\hfill\cr
}\right.
\label{Pfaffian}
\end{eqnarray}
The matrix $G(\{w\})$ is defined as follows.
If $n=$even $G(\{w\})$ is an anti-symmetric $n\times n$ matrix
with entries being fermion two-point functions:
\[
G_{ij}(\{w\})=G(w_i,w_j)\equiv
{\tr(\xi^{-2d^\phi}\phi(w_i)\phi(w_j))\ov\tr(\xi^{-2d^\phi})},
\quad 1\le i<j\le n
\]
where
\begin{eqnarray}
G^{NS}(w_1,w_2)
&=&{1\over q+q^{-1}}\sum_{m\in\Z+{1\ov 2}}\Big({w_2\ov w_1}\Big)^m
{q^{2m}+q^{-2m}\ov 1+\xi^{2m}}
\nonumber\\
&=&{1\ov 1+q^2}{\p{\xi^2;\xi^2}^3\Theta_{\xi^2}(q^4)\ov
\Theta_{\xi^2}(q^2)\Theta_{\xi^2}(-\xi q^2)}\cdot
\Big({w_2\ov w_1}\Big)^{{1\ov 2}}
{\Theta_{\xi^2}({w_2\ov w_1})\Theta_{\xi^2}(-\xi {w_2\ov w_1})\ov
\Theta_{\xi^2}(q^2{w_2\ov w_1})\Theta_{\xi^2}(q^{-2}{w_2\ov w_1})};
\nonumber\\
G^{R}(w_1,w_2)
&=&{1\over q+q^{-1}}\sum_{m\in\Z}\Big({w_2\ov w_1}\Big)^m
{q^{2m}+q^{-2m}\ov 1+\xi^{2m}}
\nonumber\\
&=&{1\ov q+q^{-1}}{\p{\xi^2;\xi^2}^3\Theta_{\xi^2}(q^4)\ov
\Theta_{\xi^2}(q^2)\Theta_{\xi^2}(-q^2)}\cdot
{\Theta_{\xi^2}({w_2\ov w_1})\Theta_{\xi^2}(-{w_2\ov w_1})\ov
\Theta_{\xi^2}(q^2{w_2\ov w_1})\Theta_{\xi^2}(q^{-2}{w_2\ov w_1})}
\nonumber\\
&=&{1\ov q+q^{-1}}{\p{\xi^2;\xi^2}^3\Theta_{\xi^2}(q^4)\ov
\Theta_{\xi^4}(q^4)}\cdot
{\Theta_{\xi^4}(({w_2\ov w_1})^2)\ov
\Theta_{\xi^2}(q^2{w_2\ov w_1})\Theta_{\xi^2}(q^{-2}{w_2\ov w_1})}
\label{propagator}
\end{eqnarray}
for Neveu-Schwarz, Ramond fermion, respectively;
the expressions are defined in a region
$|q^{-2}\xi^2|<|w_2/w_1|<|q^2|$;
here
we have introduced a theta function
\[
\Theta_x(z)
\equiv \p{z;x}\p{xz^{-1};x}\p{x;x}
=\sum_{n\in\Z}(-1)^{n}x^{{1\over2}n(n-1)}z^{n}.
\]
If $n=$odd and $\phi=\phi^R$
the matrix $G(\{w\})$ is an anti-symmetric $(n+1)\times (n+1)$ defined by
\[
G(\{w\})=\Bigg(G^-_{ij}(\{w\})\Bigg|
  \left.\matrix{i&\downarrow& 0,1,\ldots,n\cr
                j&\rightarrow& 0,1,\ldots,n\cr}\right.\Bigg);
\]
\begin{eqnarray*}
G^-_{ij}(\{w\})&=&G^{-}(w_i,w_j) \\
&\equiv&{1\over q+q^{-1}}\Big\{1+\sum_{m\in\Z_{\not=0}}\Big({w_j\ov w_i}\Big)^m
{q^{2m}+q^{-2m}\ov 1-\xi^{2m}}\Big\}, \quad 1\le i<j\le n; \\
G^-_{0j}(\{w\})
&=&{\tr_{\F^{(1)}_\epsilon}(\xi^{-2d^\phi}\phi_0)\over
\tr_{\F^{(1)}_\epsilon}(\xi^{-2d^\phi})}
={\epsilon\over[2]^{1/2}}{\p{\xi^2;\xi^2}\over\p{-\xi^2;\xi^2}},
\quad 1\le j\le n. \\
\end{eqnarray*}
For derivations see Appendix.

Note that the two-point function $G(w_1,w_2)$ becomes the delta function
$\delta(z) = \sum_{m\in\Z} z^{m}$ at $\xi=q^2$:
\begin{eqnarray}
G^{NS,R}(w_1,w_2)\Big|_{\xi=q^2}
&=&{1\over q+q^{-1}}
\Big(q^{-2}{w_2\over w_1}\Big)^{i/2}\delta\Big(q^{-2}{w_2\over w_1}\Big).
\label{gets-delta}\\
&&
(i=1,0\;{\rm according\; to}\;NS,R)
\nonumber
\end{eqnarray}

\subsection{Integral formula}
\label{sec-formula-formula}
In order to describe formulae concisely
let us introduce some notations and an order `$<$':
for the set of indices $I \equiv \{i_1,\ldots,i_n\}$,
which indicate the components of vertex operators in the product
$\Phi_{i_1}(z_1)\cdots\Phi_{i_n}(z_n)$,
let
\begin{equation}
I_j=\{l(1\le l\le n)|i_l=j\},\quad s_j=\#I_j \quad (j=0,1,2);
\end{equation}
note that by definition $s_0+s_1+s_2=n$;
let
\begin{equation}
Z=\{z_l(1\le l\le n)\},\quad
W=\{w_1^{(l)}(l\in I_1),\;w_1^{(l)},w_2^{(l)}(l\in I_0)\}
\end{equation}
where
$w_1^{(l)}$ is the integral variable (dummy) of $\Phi_{1}(z_l), l\in I_1$,
and
$w_1^{(l)},w_2^{(l)}$ the integral variables of $\Phi_{0}(z_l), l\in I_0$
(cf. (\ref{VO}));
in a set of variables $Z\cup W$
define an order `$<$' by:
(i) if $l<l'$  then
\begin{equation}
z_l<z_{l'},\; w_i^{(l)}<w_j^{(l')},\; z_l<w_j^{(l')},\; w_i^{(l)}<z_{l'};
\end{equation}
(ii) if $l=l'$ then
\begin{equation}
z_l<w_j^{(l)},\; w_1^{(l)}<w_2^{(l)}({\rm if}\;l\in I_0);
\end{equation}
we further define $W_l\equiv\{w\in W|z_l<w<z_{l+1}\}$.


\bigskip
{\em Remark.}
The number of the fermion fields $\phi(w)$ and
the number of the $w$-integrals in the product
$\Phi_{i_1}(z_1)\cdots\Phi_{i_n}(z_n)$
are $2s_0+s_1$.
Furthermore,
for the nontrivial (nonzero) trace
we have $s_0=s_2$ and, therefore, $n=2s_0+s_1$.
\bigskip

After tedious but straightforward computations
we obtain an integral formula
for the $n$-point function of vertex operators (\ref{def-bF}):
\begin{eqnarray}
&&\bF^{(0)}_{i_1,\ldots,i_n}(z_1,\ldots,z_n|\xi,y)
\nonumber\\
&&\qquad
=a_I(Z|\xi,y)
\prod_{w\in W}\oint{dw\ov 2\pi i}\;
B_I(W,Z|y)C^{(0)}_I(W,Z|\xi,y),
\label{integral-formula-0}\\
&&\bF^{(1,\ep)}_{i_1,\ldots,i_n}(z_1,\ldots,z_n|\xi,y)
\nonumber\\
&&\qquad
=a_I(Z|\xi,y)
\prod_{w\in W}\oint{dw\ov 2\pi i}\;
B_I(W,Z|y)C^{(1,\ep)}_I(W,Z|\xi,y),
\label{integral-formula-1}
\end{eqnarray}
where
\begin{eqnarray*}
&&a_I(Z|\xi,y)
\nonumber\\
&&\qquad
=(-1)^{s_1+\sum_{l=2}^n(l-1)(i_l-1)}[2]^{s_1+{1\ov 2}n}
q^{-7s_0-3s_1+4\sum_{l=2}^n(l-1)(i_l-1)}
{f(1|\xi^2)^{n}\ov\p{\xi^2;\xi^2}}
\nonumber\\
&&\quad\qquad\times
\prod_{l=1}^{n-1}z_l^{\sum_{l'>l}(i_{l'}-1)}\prod_{l\in I_1}{1\ov z_l}
\prod_{l\in I_0}{1\ov z_l^2} \prod_{z<z'}f({z'\ov z}|\xi^2),
\nonumber\\
&&B_I(W,Z|\xi)
\nonumber\\
&&\qquad
=\prod_{l=1}^{n-1}\Bigg\{\prod_{w\in W_l}w^{-\sum_{l'>l}(i_{l'}-1)}\Bigg\}
\prod_{l\in I_1}{1\ov w_1^{(l)}}\prod_{l\in I_0}{1\ov w_2^{(l)}}
\nonumber\\
&&\quad\qquad\times
{\prod_{w<w'}f({w'\ov w}|\xi^2)
\ov \prod_{z<w}f({w\ov q^4z}|\xi^2)\prod_{w<z}f({q^4z\ov w}|\xi^2)},
\nonumber\\
&&C^{(0)}_I(W,Z|\xi,y)
\nonumber\\
&&\qquad
=\tr \Big(\xi^{-2d^\phi}
  {\buildrel\to\over{\prod_{w}}}\phi^{NS}(w) \Big)\cdot
\prod_w w^{{1\ov 2}}\sum_{n\in\Z}\xi^{n^2}
\Bigg(y^2{\prod_z(-q^4z)\ov\prod_ww}\Bigg)^{n},
\nonumber\\
&&C^{(1,\ep)}_I(W,Z|\xi,y)
\nonumber\\
&&\qquad
=\tr \Big(\xi^{-2d^\phi}
  {\buildrel\to\over{\prod_{w}}}\phi^{R}(w) \Big)\cdot
\prod_w w^{{1\ov 2}}\sum_{n\in\Z+{1\over2}}\xi^{n^2}
\Bigg(y^2{\prod_z(-q^4z)\ov\prod_ww}\Bigg)^{n},
\end{eqnarray*}
and
\begin{eqnarray*}
&&f(z|x)\equiv (q^2z;x)_\infty(q^2xz^{-1};x)_\infty.
\end{eqnarray*}
The contours of $w$-integrals are anti-clockwise around the origin
such that
\[
|\xi^2q^6z|<|w|<|q^2z|\;\;{\rm for}\; z<w, \quad
|q^6z|<|w|<|\xi^{-2}q^2z|\;\;{\rm for}\; z>w;
\]
and, additionally,
\[
|\xi^2q^{-2}|<|{w'\ov w}|<|q^2|
\]
must be forced
if the integrand contains fermion two-point functions $G(w,w')$ ($w<w'$)
when expanding the fermion Pfaffian $G(\{w\})$.
The fermion traces are given in \ref{sec-formula-traces}
and depend on the order in the set $W$.

Many of the factors
$f({w'\over w}|\xi^2)$
in the numerator of the integrand
are cancelled
by those in the denominators of the fermion two-point function $G(w,w')$
(cf. Eq.(\ref{propagator})).
Observe that the contours get pinched when letting
$\xi\to q^2$ with $|\xi|<|q^2|$.
We regard the formulae as analytic functions of $\xi$,
and define them outside the region $|\xi|<|q^2|$ by analytic continuation;
we shall demonstrate it in the next subsection \ref{sec-formula-example}.
As noted before, just at $\xi=q^2$,
the fermion trace yields the delta function (see (\ref{gets-delta})),
and the formula gets simplified;
this point is further discussed in \ref{sec-simplified-simplified}.
A specialized formula at $\xi=q^2$ is important in an application to
spin correlation functions of an $S=1$ quantum spin chain,
which will be discussed in \ref{sec-simplified-spin}.

Finally we note that
$a_I(Z|\xi,y)$ and $B_I(W,Z|\xi)$ are even functions of $\xi$.

\subsection{Example: VO two-point functions}
\label{sec-formula-example}
To illustrate the content of the formula
(\ref{integral-formula-0}), (\ref{integral-formula-1}),
we write down all two-point functions of vertex operators:
\begin{eqnarray*}
&&\bF^{(0)}_{20}(z_1,z_2|\xi,y)
=-[2]q^{-11}{f(1|\xi^2)^2\over\p{\xi^2;\xi^2}}
{f({z_2\over z_1}|\xi^2)\over z_1z_2^2} \\
&&\times
\oi1\oi2
{w_1^{{1\over2}}w_2^{-{1\over2}}
f({w_2\over w_1}|\xi^2)G^{NS}(w_1,w_2)\p{-\xi;\xi^2}
\sum_{n\in\Z}\xi^{n^2}(y^2{q^8z_1z_2\over w_1w_2})^n
\over
\prod_{i,j=1,2}f({w_i\over q^4z_j}|\xi^2)}, \\
&&\bF^{(0)}_{02}(z_1,z_2|\xi,y)
=-[2]q^{-3}{f(1|\xi^2)^2\over\p{\xi^2;\xi^2}}
{f({z_2\over z_1}|\xi^2)\over z_1} \\
&&\times
\oi1\oi2
{w_1^{-{1\over2}}w_2^{-{3\over2}}
f({w_2\over w_1}|\xi^2)G^{NS}(w_1,w_2)\p{-\xi;\xi^2}
\sum_{n\in\Z}\xi^{n^2}(y^2{q^8z_1z_2\over w_1w_2})^n
\over
\prod_{i=1,2}\{f({w_i\over q^4z_1}|\xi^2)
               f({q^4z_2\over w_i}|\xi^2)\}}, \\
&&\bF^{(0)}_{11}(z_1,z_2|\xi,y)
=[2]^3q^{-6}{f(1|\xi^2)^2\over\p{\xi^2;\xi^2}}
{f({z_2\over z_1}|\xi^2)\over z_1z_2} \\
&&\times
\oi1\oi2
{w_1^{-{1\over2}}w_2^{-{1\over2}}
f({w_2\over w_1}|\xi^2)G^{NS}(w_1,w_2)\p{-\xi;\xi^2}
\sum_{n\in\Z}\xi^{n^2}(y^2{q^8z_1z_2\over w_1w_2})^n
\over
f({w_1\over q^4z_1}|\xi^2)f({w_2\over q^4z_1}|\xi^2)f({w_2\over q^4z_2}|\xi^2)
f({q^4z_2\over w_1}|\xi^2)};
\end{eqnarray*}
$\bF^{(1,\ep)}_{ij}$ are obtained from $\bF^{(0)}_{ij}$
by the replacement
\[
G^{NS}(w_1,w_2)\p{-\xi;\xi^2}\sum_{n\in\Z}
\quad\mapsto\quad
G^{R}(w_1,w_2)\p{-\xi^2;\xi^2}\sum_{n\in\Z+{1\over 2}}.
\]
The contours are restricted in the regions described below
Eqs.(\ref{integral-formula-0}) and (\ref{integral-formula-1}):
for example,
$|\xi^2q^6z_i|<|w_j|<|q^2z_i|$ ($i,j=1,2$),
$|\xi^2q^{-2}|<|w_2/w_1|<|q^2|$ for $\bF^{(0)}_{20}$, etc.

Let us examine the two-point function $\bF^{(0)}_{20}(z_1,z_2|\xi,y)$
as a function of $\xi$ more closely
and show how to obtain its analytic continuations.
The factor $f({w_2\ov w_1}|\xi^2)$ in the integrand is cancelled by
that from $G^{NS}$.
Apart from poles at
$\xi^{2n}q^6z_i$, $\xi^{-2n}q^2z_i$ ($n=0,1,\ldots$; $i=1,2$),
the integrand as a function of $w_2$ has poles at
\[
\ldots,\; \xi^6q^{-2}w_1,\; \xi^4q^{-2}w_1,\; \xi^2q^{-2}w_1;\;
q^2w_1,\; \xi^{-2}q^{2}w_1,\; \xi^{-4}q^{2}w_1,\; \ldots.
\]
The original contour of the $w_2$-integral is confined in a region between
the poles $\xi^2q^{-2}w_1$ and $q^2w_1$.
We deform it to a sum of
a contour anti-clockwise around the origin
in a region $|q^2w_1|<|w_2|<|\xi^{-2}q^2w_1|$
and
a one clockwise around the pole at $w_2=q^2w_1$
(which counts minus a residue),
so as to get an expression of $\bF_{20}^{(0)}$ as a function of $\xi$
in a neighbourhood of $\xi=q^2$.
Noting that
\begin{eqnarray*}
&&\bigg[f\Big({w_2\over w_1}\Big|\xi^2\Big)
G^{NS,R}(w_1,w_2)\cdot (w_2-q^2w_1)\bigg]_{w_2\to q^2w_1} \\
&&\qquad\qquad\qquad={-q^{2}\ov [2]}\p{q^4;\xi^2}\p{\xi^2;\xi^2}w_1
\end{eqnarray*}
the residue at $w_2=q^2w_1$ is evaluated simply;
the result is
\begin{eqnarray}
\bF_{20}^{(0)}(z_1,z_2|\xi,y)
&=&-q^{-10}\p{q^4;\xi^2}f(1|\xi^2)^2{f({z_2\ov z_1}|\xi^2)\ov z_1z_2^2}
\nonumber\\
&&\times
\oi1{w_1\p{-\xi;\xi^2}\sum_{n\in\Z}\xi^{n^2}\Big(y^2{q^6z_1z_2\ov w_1^2}\Big)^n
\ov
\prod_{i=1,2}\{f({w_1\ov q^4z_i}|\xi^2)f({w_1\ov q^2z_i}|\xi^2)\}
}
\nonumber\\
&&+({\rm double\;integral\;with\;}|q^2|<|{w_2\ov w_1}|<|\xi^{-2}q^2|).
\label{analytic-continuation}
\end{eqnarray}
This is an analytic continuation of the original formula to a neighborhood
of $\xi=q^2$.
Observe that
the second term vanishes at $\xi=q^2$,
since it includes a factor $\Theta_{\xi^2}(q^4)$
which arose from the fermion two-point function;
this fact follows directly from (\ref{gets-delta}).

\section{Simplification at $\xi=q^2$ and an application to a spin chain}
\label{sec-simplified}
We have seen that the formula for the VO two-point functions
gets simplified
when specialized at $\xi=q^2$ (Section \ref{sec-formula-example}).
In this section
we first write down all simplified integral formula for
the VO two-point functions at $\xi=q^2$.
Then we describe how to obtain
a simplified integral formula for the VO $n$-point function at $\xi=q^2$
from the general formulae
(\ref{integral-formula-0}) and (\ref{integral-formula-1}).
In this simplified formula
the number of contour integrals reduces from $n$ to $n/2$.
It is due to a property of the fermion two-point function
which becomes the delta function at $\xi=q^2$.
An application to the $S=1$ spin chain is given in \ref{sec-simplified-spin}.

We restrict ourselves to $n=$even.

\subsection{VO two-point functions at $\xi=q^2$}
\label{sec-simplified-example}
{}From the first term of the equation (\ref{analytic-continuation})
we get an expression of $\bF^{(0)}_{20}$ at $\xi=q^2$;
similarly we get other VO two-point functions at $\xi=q^2$:
\begin{eqnarray*}
&&\bF^{(0)}_{20}(z_1,z_2|q^2,y)
=-q^{-10}\p{q^4;q^4}f(1|q^4)^2{f({z_2\over z_1}|q^4)\over z_1z_2^2}
\nonumber\\
&&\quad\times
\oi1 {w_1\p{-\xi;q^4}\sum_{n\in\Z}\xi^{n^2}(y^2{q^6z_1z_2\over w_1^2})^n
\over
\p{q^{-2}{w_1\over z_1};q^2}\p{q^{-2}{w_1\over z_2};q^2}
\p{q^{8}{z_1\over w_1};q^2}\p{q^{8}{z_2\over w_1};q^2} },
\nonumber\\
&&\bF^{(0)}_{02}(z_1,z_2|q^2,y)
=-q^{-4}\p{q^4;q^4}f(1|q^4)^2{f({z_2\over z_1}|q^4)\over z_1}
\nonumber\\
&&\quad\times
\oi1 {w_1^{-1}\p{-\xi;q^4}\sum_{n\in\Z}\xi^{n^2}(y^2{q^6z_1z_2\over w_1^2})^n
\over
\p{q^{-2}{w_1\over z_1};q^2}\p{q^{2}{w_1\over z_2};q^2}
\p{q^{8}{z_1\over w_1};q^2}\p{q^{4}{z_2\over w_1};q^2} },
\nonumber\\
&&\bF^{(0)}_{11}(z_1,z_2|q^2,y)
=[2]^2q^{-5}\p{q^4;q^4}f(1|q^4)^2{f({z_2\over z_1}|q^4)\over z_1z_2}
\nonumber\\
&&\quad\times
\oi1 {\p{-\xi;q^4}\sum_{n\in\Z}\xi^{n^2}(y^2{q^6z_1z_2\over w_1^2})^n
\over
\p{q^{-2}{w_1\over z_1};q^2}\p{{w_1\over z_2};q^2}
\p{q^{6}{z_1\over w_1};q^2}\p{q^{4}{z_2\over w_1};q^2} };
\nonumber
\end{eqnarray*}
$\bF^{(1,\ep)}_{ij}$ are obtained from $\bF^{(0)}_{ij}$
by the replacement
\[
\p{-\xi;q^4}\sum_{n\in\Z}
\quad\mapsto\quad
\p{-\xi^2;q^4}\sum_{n\in\Z+{1\over 2}}.
\]
We have reserved odd $\xi$ for convenience of (anti-)symmetrization
with respect to $\xi$,
which is necessary to obtain expressions of $\widetilde F^{(\lambda)}_{ij}$
(cf. Section \ref{sec-formula-def}).

\subsection{Simplified formula for the VO $n$-point function at $\xi=q^2$}
\label{sec-simplified-simplified}
We have seen that the formula for the VO two-point functions
gets greatly simplified
when specialized at $\xi=q^2$.
This feature holds as well for the VO $n$-point functions;
the number of integrals reduces to ${1\ov 2}n$ at $\xi=q^2$.


A simplified formula for the VO $n$-point function
at $\xi=q^2$ is obtained as follows:
first we expand the Pfaffian in the formula (\ref{integral-formula-0})
or (\ref{integral-formula-1}) into $(n-1)(n-3)\cdots3\cdot1$ terms,
each of which being a product of $n/2$ fermion two-point functions,
say,
\[
G(w_{r_{1}},w_{r_{2}})G(w_{r_{3}},w_{r_{4}})\cdots G(w_{r_{n-1}},w_{r_{n}});
\]
deform the contours of $w_{r_{2}},w_{r_{4}},\ldots,w_{r_{n}}$
and evaluate minus the residue at
$q^2w_{r_{1}},q^2w_{r_{3}},\ldots,q^2w_{r_{n-1}}$, respectively,
as in the case of two-point functions in the previous subsection;
let $\xi\to q^2$;
then we get an formula containing only
$w_{r_{1}},w_{r_{3}},\ldots,w_{r_{n-1}}$ contour integrals
(the rest terms with more integrals vanish).

Practically
we are to replace
\[
f\Big({w_{r_{2j}}\over w_{r_{2j-1}}}\Big|\xi^2\Big)
G(w_{r_{2j-1}},w_{r_{2j}})
\longmapsto
{q^2\over [2]}\p{q^4;\xi^2}\p{\xi^2;\xi^2}w_{r_{2j-1}}
\]
for all $j=1,2,\ldots,{n\over2}$,
remove $\oint{dw_r\over2\pi i}$, $r=r_2,r_4,\ldots,r_n$,
and put $\xi=q^2$,
$w_{r_{2j}}=q^2w_{r_{2j-1}}, j=1,2,\ldots,{n\over2}$.
Then we obtain a simplified formula at $\xi=q^2$.

\subsection{Application to an integrable $S=1$ spin chain}
\label{sec-simplified-spin}
The spin $n$-point function for the integrable $S=1$ spin chain
(the spin-1 analog of the XXZ model) is given by the formula (\ref{spin-corr})
with $k=2$;
note that $\p{q^2;q^4}/\p{q^6;q^4} = 1-q^2$.

The formula can be written in terms of $\bF$:
\begin{eqnarray}
&&\br{E_{i_nj_n}\otimes\cdots\otimes E_{i_1j_1}}^{(\lambda_m)}_{z_n,\ldots,z_1}
=\epsilon^{mn}(-1)^{mn+\sum_{l=1}^{n}(i_l-1)}
\nonumber\\
&&\times
q^{2n+\sum_{l=1}^{n}(2-i_l)(1-i_l)}
(1-q^2)^n
[2]^{-\#\{1\le l\le n|i_l=1\}}
{(q^2)^{-{1\over2}(\lambda_m,\lambda_m)}
\over
\tr_{V(\lambda_m)}(q^{-2\rho})}
\nonumber\\
&&\times
\prod_{l=1}^{n}z_l\cdot
\bF^{(*)}_{k-i_1,\ldots,k-i_n,j_n,\ldots,j_1}
(z_1q^{-2},\ldots,z_nq^{-2},z_n,\ldots,z_1|q^2,q^{-1})
\label{spin-corr-k=2}
\end{eqnarray}
where $\bF^{(*)}$ stands for
$[\bF^{(0)}]^{\sigma_0}$,
$[\bF^{(0)}]^{\sigma_1}$,
$\bF^{(1,\epsilon)}$ for $m=0,2,1$, respectively.
$\rho = \Lambda_0+\Lambda_1 = 2d+{1\over 2}\alpha$ (in our representation),
$\lambda_m = (2-m)\Lambda_0+m\Lambda_1$,
$(\quad,\quad)$ being given in (\ref{inner-product}).
(This formula for $m=1$ appears to depend on our choice of realization
$V(\Lambda_0+\Lambda_1) = \F^{(1)}_\epsilon $, $\epsilon = \pm1$;
but it does not actually as can be seen if we substitute
(\ref{integral-formula-1}).)

For the spin chain correlations we set $z_1=\cdots z_n$;
equation (\ref{spin-corr-k=2}) itself with general $(z_1,\ldots,z_n)$
is the correlation function of the inhomogeneous vertex model.
The range of the parameter $q$ is $-1<q<0$;
but it might be extended to $|q|<1$
(cf. Section \ref{sec-spin-corr-general}).

As an example we consider the spin one-point functions,
which are related to VO two-point functions, and represented by
a single contour integral respectively.
It is easily seen that identities
$\br{E_{jj}}^{(\lambda_{m})} = \br{E_{2-j,2-j}}^{(\lambda_{2-m})},j=0,1,2$,
hold.
If we expand the expressions in $q$ by hand
we get
\begin{eqnarray*}
&&\br{E_{22}}^{(2\Lambda_0)}_z
=1-2q^2+5q^6-2q^8-9q^{10}+3q^{12}+19q^{14}+O(q^{16}),\\
&&\br{E_{00}}^{(2\Lambda_0)}_z
=2q^4+q^6-4q^8-9q^{10}+7q^{12}+19q^{14}+O(q^{16}),\\
&&\br{E_{11}}^{(2\Lambda_0)}_z
=2q^2-2q^4-6q^6+6q^8+18q^{10}-10q^{12}-38q^{14}+O(q^{16}),\\
&&\br{E_{00}}^{(\Lambda_0+\Lambda_1)}_z
=2q^2-4q^4+2q^6+6q^{10}-4q^{12}-6q^{14}+O(q^{16}),\\
&&\br{E_{11}}^{(\Lambda_0+\Lambda_1)}_z
=1-4q^2+8q^4-4q^6-12q^{10}+8q^{12}+12q^{14}+O(q^{16}).
\end{eqnarray*}
{}From this list we have
\[
\br{s^z}^{(2\Lambda_1)}_z=1-2q^2-2q^4+4q^6+2q^8-4q^{12}+O(q^{16})
\]
and $\br{s^z}^{(\Lambda_0+\Lambda_1)}_z=0$;
the both agree with the Bethe Ansatz results in Ref.\cite{DJMO}.
We note that the equation for
$\br{s^z}^{(2\Lambda_1)}_z$
in Ref.\cite{DJMO} can be factorized as
$\br{s^z}^{(2\Lambda_1)}_z=\p{q^2;q^2}^2/\p{-q^4;q^4}$.

\section*{Acknowledgement}
This paper is a thoroughly revised and rewritten version of my previous
manuscript \cite{preprint} and the thesis \cite{Thesis} in which,
compared with the present one,
I emphasized an application to the spin chain correlations,
in particular, spin one-point functions,
and did not write down the integral formula for $n$-point functions.
I wish to thank Michio Jimbo for suggesting me that my previous
results can be simplified more, and for discussions from which I learned much.
I also thank T. Miwa and T. Tokihiro for discussions and assistances
in completing the previous work.

After completing the previous version \cite{preprint,Thesis}
there appeared a preprint \cite{Weston} in which they considered
the same spin chain correlations and gave an integral formula for them;
but they did not give the construction of $V(\Lambda_0+\Lambda_1)$
and their formula in \cite{Weston} seems more complicated than ours
in the present paper;
I am grateful to Robert A. Weston for communications and his interest in
my work \cite{preprint,Thesis}.

\section*{Appendix: Boson fermion calculus}
In this Appendix we assemble
formulae of normal ordering and traces
needed for deriving or proving formulae in the text.
Among other things we must be careful with the Ramond fermion.


\subsection*{A.1  Boson calculus}

\subsubsection*{Normal ordering}
A normal order product $:A:$ of an element $A$ of the boson algebra
is defined as usual:
annihilation operators are placed at the right to the creation operators;
for example, $:a_ma_{-n}: =a_{-n}a_m$ ($m,n>0$).

We collect here formulae for the normal ordering of exponentials of
boson operators which are used in this paper;
recall that
\begin{eqnarray*}
\Phi_2(z)
&=&F_<(z)F_>(z)e^{\alpha} (-q^4z)^{{1\ov 2}\partial_\alpha}, \\
x^+(z)
&=&[2]^{{1\ov 2}}E_<^+(z)E_>^+(z)\phi(z)
e^\alpha z^{{1\over 2}+{1\over 2}\partial_\alpha}, \\
x^{-}(z)
&=&[2]^{{1\ov 2}}E_<^-(z)E_>^-(z)\phi(z)
e^{-\alpha} z^{{1\over 2}-{1\over 2}\partial_\alpha},
\end{eqnarray*}
where
\begin{eqnarray*}
&&F_<(z)=
\exp\big(\sum_{m=1}^\infty{a_{-m}\over[2m]}q^{5m}z^m\big),\quad
F_>(z)=
\exp\big(-\sum_{m=1}^\infty{a_{m}\over[2m]}q^{-3m}z^{-m}\big),\\
&&E_<^+(z)=
\exp\big(\sum_{m=1}^\infty{a_{-m}\over[2m]}q^{-m}z^m\big),\quad
E_>^+(z)=
\exp\big(-\sum_{m=1}^\infty{a_{m}\over[2m]}q^{-m}z^{-m}\big),\\
&&E_<^-(z)=
\exp\big(-\sum_{m=1}^\infty{a_{-m}\over[2m]}q^{m}z^m\big),\quad
E_>^-(z)=
\exp\big(\sum_{m=1}^\infty{a_{m}\over[2m]}q^{m}z^{-m}\big);
\end{eqnarray*}
we have the following:
\begin{eqnarray*}
F_>(z_1)F_<(z_2)&=&(1-q^{2}z_2/z_1)F_<(z_2)F_>(z_1), \\
F_>(z_1)E^-_<(z_2)&=&{1\ov 1-q^{-2}z_2/z_1}E^-_<(z_2)F_>(z_1), \\
E^-_>(z_1)F_<(z_2)&=&{1\ov 1-q^{6}z_2/z_1}F_<(z_2)E^-_>(z_1), \\
E^-_>(z_1)E^-_<(z_2)&=&(1-q^{2}z_2/z_1)E^-_<(z_2)E^-_>(z_1), \\
&&\\
F_>(z_1)E^+_<(z_2)&=&(1-q^{-4}z_2/z_1)E^+_<(z_2)F_>(z_1), \\
E^+_>(z_1)F_<(z_2)&=&(1-q^{4}z_2/z_1)F_<(z_2)E^+_>(z_1), \\
E^+_>(z_1)E^+_<(z_2)&=&(1-q^{-2}z_2/z_1)E^+_<(z_2)E^+_>(z_1), \\
&&\\
E^+_>(z_1)E^-_<(z_2)&=&{1\ov 1-z_2/z_1}E^-_<(z_2)E^+_>(z_1), \\
E^-_>(z_1)E^+_<(z_2)&=&{1\ov 1-z_2/z_1}E^+_<(z_2)E^-_>(z_1);
\end{eqnarray*}
the right hand sides are normal order products.

\subsubsection*{Traces of boson exponentials}
\begin{eqnarray*}
&&\tr_{\F^a}(x^{\sum_{m>0}mN_m^a}\exp(-\sum_{m>0}a_{-m}f_m)\exp(\sum_{m>0}a_{m}f_{-m})) \\
&&\quad=\prod_{m=1}^\infty{1\ov
1-x^m}\times\prod_{m=1}^\infty\exp\big(-[a_m,a_{-m}]f_mf_{-m}{x^m\ov
1-x^m}\big).
\end{eqnarray*}
We have used
\[
\sum_{n=k}^\infty{n!\ov k!(n-k)!}X^n = {1\ov 1-X}\big({X\ov 1-X}\big)^k.
\]
Further specialization
$[a_m,a_{-m}]=[2m]^2/m$, $A_m={q^m\ov [2m]}f_m$, $A_{-m}={q^m\ov [2m]}f_{-m}$
and $f_mf_{-m}=\sum_y\ep_yy^m$
yields
\begin{eqnarray*}
&&\tr_{\F^a}(x^{\sum_{m>0}mN_m^a}\exp(-\sum_{m>0}a_{-m}f_m)\exp(\sum_{m>0}a_{m}f_{-m})) \\
&&\quad={1\ov \p{x;x}}\times\prod_y\p{q^2xy;x}^{\ep_y}.
\end{eqnarray*}

\subsection*{A.2  Neveu-Schwarz fermion}

\subsubsection*{Operator product expansions}
Operator product expansions of two and three Neveu-Schwarz fermion fields
are
\begin{eqnarray}
\phi(w_1)\phi(w_2)
&=&\br{\phi(w_1)\phi(w_2)}
+ :\phi(w_1)\phi(w_2):,
\label{OPE2}\\
\phi(w_1)\phi(w_2)\phi(w_3)
&=&
\br{\phi(w_1)\phi(w_2)}\phi(w_3)
- \br{\phi(w_1)\phi(w_3)}\phi(w_2)
\nonumber\\
&&+ \br{\phi(w_2)\phi(w_3)}\phi(w_1)
+ :\phi(w_1)\phi(w_2)\phi(w_3):
\label{OPE3}
\end{eqnarray}
where
the singular part is
\begin{eqnarray*}
\br{\phi(w_1)\phi(w_2)}
&=&{({w_2\ov w_1})^{1/2}(1-{w_2\ov w_1})\ov
(1-q^2{w_2\ov w_1})(1-q^{-2}{w_2\ov w_1})}
\end{eqnarray*}
and the product $\phi(w_1)\phi(w_2)$ is defined in a region
$|q^2{w_2\ov w_1}|,|q^{-2}{w_2\ov w_1}|<1$.

The normal order products are defined as usual by
\begin{eqnarray*}
:\phi(w_1)\phi(w_2):
&=&\sum_{m,n} {\cal N}[\phi_m\phi_n] w_1^{-m}w_2^{-n}, \\
:\phi(w_1)\phi(w_2)\phi(w_3):
&=&\sum_{m,n} {\cal N}[\phi_m\phi_n\phi_l] w_1^{-m}w_2^{-n}w_3^{-l}
\end{eqnarray*}
where ${\cal N}$ is defined for two-products by
\begin{eqnarray*}
{\cal N}[\phi_m\phi_n]
&=&-\phi_n\phi_m \quad {\rm if}\;m>0,n<0 \\
&=&\phi_m\phi_n \quad {\rm otherwise}
\end{eqnarray*}
and similarly for three-products;
note that they are totally anti-symmetric: for example,
$:\phi(w_1)\phi(w_2): = -:\phi(w_2)\phi(w_1):$.

\subsubsection*{Traces:  Wick's theorem}
The $n$-point function of fermion fields is written by a Pfaffian
of a matrix whose entries are two-point functions (Wick's theorem).

Introduce a notation
\[
\bbr{A}
\equiv{\tr_{\F^\phi}(\xi^{-2d^\phi}A)\ov\tr_{\F^\phi}(\xi^{-2d^\phi})}.
\]
Noting that
$[\phi_{\pm m},N_m^\phi]=\pm \phi_{\pm m}(m>0)$,
$\xi^{2d^\phi}\phi_{\pm m}\xi^{-2d^\phi}=\xi^{\pm 2m}\phi_{\pm m}(m>0)$,
we can derive the following:
\begin{eqnarray*}
\bbr{\phi(w_1)\phi(w_2)}
&=&{1\ov q+q^{-1}}\sum_{m\in\Z+{1\ov 2}}\big({w_2\ov w_1}\big)^m
{q^{2m}+q^{-2m}\ov 1+\xi^{2m}},
\end{eqnarray*}
\begin{eqnarray*}
\bbr{\phi(w_1)\cdots\phi(w_n)}
&=&
\left\{\matrix{
\Pf G\quad{\rm if}\;n={\rm even};\cr
0    \quad{\rm if}\;n={\rm odd},\cr
}\right.
\end{eqnarray*}
where a $n\times n$  anti-symmetric matrix $G=(G_{ij})$ is defined by
$G_{ij} = G(w_i,w_j) \equiv \bbr{\phi(w_i)\phi(w_j)}$ ($i<j$).

\subsection*{A.3  Ramond fermion}

\subsubsection*{Operator product expansions}
It seems convenient to
define the singular part and the normal order product
(artificially) by
\begin{eqnarray*}
\br{\phi(w_1)\phi(w_2)}
&=&
{1\ov [2]}{(1+{w_2\ov w_1})(1-{w_2\ov w_1})\ov
(1-q^2{w_2\ov w_1})(1-q^{-2}{w_2\ov w_1})},
\end{eqnarray*}
\begin{eqnarray*}
:\phi(w_1)\phi(w_2):
&=&\sum_{m,n} {\cal N}[\phi_m\phi_n] w_1^{-m}w_2^{-n} - (\phi_0)^2, \\
:\phi(w_1)\phi(w_2)\phi(w_3):
&=&{\cal N}[\phi(w_1)\phi(w_2)\phi(w_3)] \\
&&-(\phi_0)^2[\phi(w_1)-\phi(w_2)+\phi(w_3)]
\end{eqnarray*}
so that the normal order products are totally anti-symmetric:
for example, we have
$:\phi(w_1)\phi(w_2): = -:\phi(w_2)\phi(w_1):$.

Operator product expansions are the same as for the NS fermion,
given by (\ref{OPE2}), (\ref{OPE3}).

\subsubsection*{Traces:  Wick's theorem}
We can write the $n$-point function of the Ramond fermion fields
in terms of a Pfaffian;
in this case the function for odd $n$ does not vanish.

Define
\[
\bbr{A}
\equiv{\tr_{\F^{\phi^{R}}_\ep}(\xi^{-2d^\phi}A)\over
\tr_{\F^{\phi^{R}}_\ep}(\xi^{-2d^\phi})}
\quad
\hbox{$\F^{\phi^R}_\ep$ given in (\ref{Ramond-Fock-space})}.
\]
Direct computation yields
$\tr_{\F^{\phi^R}_\epsilon}(\xi^{-2d^\psi})=\p{-\xi^2;\xi^2}$.
We have
\[
\bbr{\phi(w_1)}
=\bbr{\phi_0} =
\psi_0 \bbr{1\otimes
\left( \begin{array}{clr}
  0 & 1 \\
  1 & 0
\end{array} \right)}
=\ep\psi_0{\p{\xi^2;\xi^2}\ov\p{-\xi^2;\xi^2}},
\]
\begin{eqnarray*}
\bbr{\phi(w_1)\phi(w_2)}
&=&{1\ov q+q^{-1}}\sum_{m\in\Z}\big({w_2\ov w_1}\big)^m
{q^{2m}+q^{-2m}\ov 1+\xi^{2m}},
\end{eqnarray*}
and in general
\begin{eqnarray*}
\bbr{\phi(w_1)\cdots\phi(w_n)}
&=&\Pf G^\pm\quad{\rm according\; to}\;n={\rm even,odd}
\end{eqnarray*}
where
if $n=$even $G^+$ is an anti-symmetric $n\times n$ matrix
with entries
$G^+_{ij}=\bbr{\phi(w_i)\phi(w_j)}$ for $i<j$;
if $n=$odd then $G^-$ is an anti-symmetric $(n+1)\times (n+1)$ matrix
\[
G^-=\Bigg(G^-_{ij}\Bigg|
  \left.\matrix{i&\downarrow& 0,1,\ldots,n\cr
                j&\rightarrow& 0,1,\ldots,n\cr}\right.\Bigg);
\]
\begin{eqnarray*}
G^-_{ij}
&=&{1\over q+q^{-1}}\Big\{1+\sum_{m\in\Z_{\not=0}}\Big({w_j\ov w_i}\Big)^m
{q^{2m}+q^{-2m}\ov 1-\xi^{2m}}\Big\}, \quad 1\le i<j\le n; \\
G^-_{0j}
&=&\bbr{\phi_0},
\quad 1\le j\le n.
\end{eqnarray*}
We note that
$[\phi_{0},N_m^\phi]=0$, $\xi^{2d^\phi}\phi_{0}\xi^{-2d^\phi}=\phi_{0}$,
and their consequence
\[
(1+(-1)^n\xi^{2m_1})\bbr{\phi_{m_1}\cdots\phi_{m_n}}
=\sum_{j=2}^{n}(-1)^j\delta_{m_1+m_j,0}[\phi_{m_1},\phi_{-m_1}]_+
\bbr{
  \overbrace{\phi_{m_2}\cdots\phi_{m_n}}^{\phi_{m_j}\;{\rm removed}}
};
\]
from this equation
we can derive the last result.



\end{document}